\documentclass[letterpaper,twocolumn,10pt]{article}
\usepackage{usenix-2020-09}

\def\mysubsection#1{\medskip\noindent{\bf #1.}} 

\newif\iffull\fulltrue 
\newif\ifdraft\draftfalse
\ifdraft
\newcommand{\cedric}[1]{\emph{\color{blue} #1}}
\else
\newcommand{\cedric}[1]{}
\fi 

\usepackage{paralist}
\usepackage{cite}
\usepackage{tikz}
\usetikzlibrary{datavisualization}
\usepackage{graphicx}
\usepackage{msc5}
\usepackage{listings}

\usepackage{color}
\definecolor{lightgray}{rgb}{.9,.9,.9}
\definecolor{darkgray}{rgb}{.4,.4,.4}
\definecolor{purple}{rgb}{0.65, 0.12, 0.82}
\lstdefinelanguage{JavaScript}{
  keywords={break, case, catch, continue, debugger, default, delete, do, else, false, finally, for, function, if, in, instanceof, new, null, return, switch, this, throw, true, try, typeof, var, void, while, with},
  morecomment=[l]{//},
  morecomment=[s]{/*}{*/},
  morestring=[b]',
  morestring=[b]",
  ndkeywords={class, export, boolean, throw, implements, import, this},
  keywordstyle=\color{blue}\bfseries,
  ndkeywordstyle=\color{darkgray}\bfseries,
  identifierstyle=\color{black},
  commentstyle=\color{purple}\ttfamily,
  stringstyle=\color{red}\ttfamily,
  sensitive=true
}

\usepackage{xcolor}
\definecolor{darkgreen}{RGB}{0,100,0}
\usepackage{hyperref}

\newcommand{\name}[1]{{\ttfamily#1}} 
\newcommand{\apex}{\name{conf}}
\begin{document}

\title{%
Transparent Attested DNS for Confidential Computing Services
\iffull
\fi 
}

\iffull
\author{ 
  {\rm Antoine Delignat-Lavaud}
  \\ \rm\normalsize Azure Research \and 
  {\rm Cédric Fournet}
  \\ \rm\normalsize Azure Research \and 
  {\rm Kapil Vaswani}
  \\ \rm\normalsize Azure Research \and 
  {\rm Manuel Costa}
  \\ \rm\normalsize Azure Research \and 
  {\rm Sylvan Clebsch}
  \\ \rm\normalsize Azure Research \and 
  {\rm Christoph M. Wintersteiger}
  \\ \rm\normalsize Imandra
}
\fi

\maketitle

\begin{abstract}
Confidential services running in hardware-protected Trusted Execution Environments (TEEs) can provide higher security assurance, but this requires custom clients and protocols to distribute and verify their attestation evidence.
Compared with classic Internet security, built upon universal abstractions such as domain names, origins, and certificates, this puts a significant burden on service users and providers.
In particular, Web browsers and other legacy clients do not get the same security guaranties as custom clients.

We present a new approach for users to establish trust in confidential services. 
We propose attested DNS (aDNS): a name service that securely binds the attested implementation of 
confidential services to their domain names.
aDNS enforces policies for all names in its zone of authority: any TEE that runs a service must present a hardware attestation that complies with the domain-specific policy before registering keys and obtaining certificates for any name in this domain.
aDNS provides protocols for zone delegation, 
TEE registration, and certificate issuance.
aDNS builds on standards such as DNSSEC, DANE, ACME and Certificate Transparency.
aDNS provides DNS transparency by keeping all records, policies, and attestations in a public append-only log, thereby enabling auditing and preventing targeted attacks.
  
We implement aDNS as a confidential service using a fault-tolerant network of TEEs. We evaluate it using sample confidential services that illustrate various TEE platforms. 
On the client side, we provide a generic browser extension that queries and verifies attestation records before opening TLS connections, with negligible performance overhead, 
and we show that, with aDNS, even legacy Web clients benefit from confidential computing as long as some enlightened clients verify attestations to deter or blame malicious actors.
\end{abstract}

\section{Introduction}

Clients routinely trust Internet services based on their domain names, such as \verb"usenix.org".
As they connect to the service (using e.g. HTTP over TLS) their networking stack ensures that the server authenticates with a valid certificate for this name, and users more generally trust that the domain owner will grant such certificates only to authorized servers. 
This conveniently hides many implementation details, such as network addresses, certificates, and server configurations, and also allows delegation of operations to third parties such as content delivery networks, caches, and other front-ends.

Confidential computing enables services to run in isolation in hardware-protected trusted execution environments (TEEs)
with increased security both for the service owner and for its users. 
TEEs that run the service present attestation reports that jointly authenticate their hardware platform, their code, and their configuration.
While this is much more informative than a domain name, 
the management, distribution, discovery, updating, and validation of attested evidence to establish the identity of a service are still largely open problems. 

To illustrate the challenges of deploying confidential computing in practice, consider 
Signal's contact discovery~\cite{signal}, one of the first successful confidential services with over 40M users. 
This directory service allows users to query the usernames associated with phone numbers. 
To protect user privacy, neither the cloud service provider (CSP) hosting the service TEEs nor the service owner should be able to learn 
these queries. 
%
%
When connecting to the service, and before querying any phone number, the Signal client checks the attestation report 
presented by the TEE to ensure it is running the correct code and configuration on an up-to-date platform.
%
%
Service owners like Signal facilitate this check 
by making attestation fine-grained and by providing the corresponding attestation-verification policy.  
Their TEE attestation includes a fresh authentication key to be used by clients to establish a secure channel to the service and encrypt their queries and responses. 
The service owner may additionally issue a certificate to endorse this attested public key.    
However, this is very brittle: 
any TEE code update or platform update requires updating clients with a new attestation-verification policy, which is hard to manage for scalable, long-lived services that involve many TEEs over their lifetime.   

Verifying attestations for a given service also requires a custom client:  
the Signal chat application code embeds specific attestation-verification policy and code, and it is unclear how they would offer the same guarantees from a Web browser visiting \name{https://signal.org}.   
From a security viewpoint, the need for service-specific clients (and more generally for trusted mechanisms to discover them and keep them up-to-date) limits the benefits of confidential computing.
Without independent scrutiny, for example, service owners could technically mount powerful targeted attacks by 
providing rogue clients that validate attestations from rogue TEEs.   

We observe that clients heavily rely on DNS to identify services today, inasmuch as X.509, HTTPS and all Web security boundaries are ultimately rooted in DNS names. 
Hence, a refinement of the Signal certificate idea above is for the TEE to turn its attested key into a certificate for a well-known service name such as \name{directory.signal.org}. 

The core novelty of this paper is to show how to 
systematically distribute attested keys 
by turning DNS itself into a confidential service called attested DNS (aDNS).
aDNS acts as a trusted controller for confidential services, i.e., it can authorize TEEs to participate in the service in a given role, 
and it can control the issuance of certificates to those TEEs.
aDNS also makes the {\em governance} of the confidential service explicit by recording 
and enforcing policies that authorize its TEEs, and by making them discoverable to clients.
%
%
Lastly, aDNS makes confidential services {\em auditable} by recording their history of policies,
attestations, and certificates. 

aDNS builds on standard security protocols such as TLS, DNSSEC, DANE, and ACME, 
which increases the security of records and certificates accepted both by enlightened clients (that verify attestations
and can blame bad actors) and legacy clients (that delegate this verification to aDNS), thereby providing a clear path for incremental deployment of confidential computing. 
Altogether, its features yield a new security foundation layer for the Web by providing a universal (i.e. service-agnostic) definition of confidential services: any service whose name is controlled by aDNS.
Our definition comes with a familiar name-based user interface, supporting statements of the form ``every service whose name ends with \name{*.conf} runs in TEEs that meet its stated service policy.''

Note that aDNS does not ascribe any semantics to the TEEs that participate in the service (in particular, it doesn't in itself guarantee that the data submitted to the service is confidential to the CSP and owners), but it enforces the intent of the service owners expressed as policies, which can be discovered and audited.
As an example, a trading website may want to provide fairness of order execution, or a source code management website may want to guarantee the consistency of contributed commits using their TEEs. These are all valid examples of {\em confidential services}.
We refer readers to the closely related problem of  {\em code transparency}~\cite{contour18,delignat2023should} to understand how to assign security properties to binary measurements of TEEs.

We implement an aDNS server and a browser extension client, and we show that they can be used to access different types of confidential services on various TEE platforms. 
We carefully optimize the connection protocol for aDNS clients such that checking attestation doesn't introduce any significant latency overhead.
We also leverage the existing distributed DNS cache infrastructure to ensure scalability of aDNS, applying many of the lessons from similar DNS-based connection-time extensions in browsers, such as the recently deployed HTTPS/SVCB records.
Our evaluation shows that the total connection latency overhead of aDNS (from the user intent to the first request packet) is negligible. 

In summary, we make the following contributions:
\begin{enumerate}
  \item a transparent attested DNS architecture to securely bind confidential service implementations to their domain names based on their registration policies (
  \S\ref{sec:arch}, \S\ref{sec:threat-model});
  \item protocols for querying and verifying attestation records, for registering TEEs as part of a service, and for obtaining certificates from an ACME-compliant certificate authority based on their attestation reports (\S\ref{sec:protocols});
  \item protocols for controlled zone delegation, for bootstrapping trust, and for integration with DNSSEC PKI (\S\ref{sec:delegation});

  \item implementations of an attested authoritative DNS service running as a network of TEEs, and of a lightweight browser extension for Web clients (\S\ref{sec:impl});
  \item sample confidential services (ML inference, token issuance, and privacy-preserving ad selection) integrated with aDNS, showing support for diverse TEEs: SGX enclaves, SEV-SNP confidential VMs, and confidential containers (\S\ref{sub:apps}); 
  \item an experimental evaluation of our implementation tested on the sample applications (\S\ref{sec:eval});  


\end{enumerate}
In addition, \S\ref{sec:background} provides background on the technologies on which aDNS is built: DNS and DNSSEC, confidential computing, and transparency ledgers; \S\ref{sec:related} discusses related work; and  
an appendix provides additional details on sample applications and attested CAs. 

%
%

\section{Background}
\label{sec:background}

\noindent{\bf DNS, DNSSEC, and DANE.}
%
%
The Domain Name System (DNS), one of the most important foundations of the Internet, allows domain names to be associated with network addresses, services, and other resources encoded as {\em resource records} (RR). 
Names range over series of labels such as \name{usenix} and \name{org}.   
The ability to delegate the authority for names under a given suffix is the most defining feature of DNS, 
and is a major reason for its success as a distributed system: 
it offers full independent control of the delegated instance while simultaneously allowing a rough form of Internet governance at the upper layers of the hierarchy, which helps mitigate some of the problems such as namesquatting, typosquatting, Sibyls, or name conflict resolution that plague other distributed identity systems~\cite{szurdi2014long, kalodner2015empirical}.
The concept of {\em (authority) zone} naturally emerges from delegation, and represents all the names controlled by a DNS instance that are not delegated.
By policy, the root zone (controlled by ICANN) only contains delegations, whose names are called top-level domains (TLDs), such as .uk (delegated to Nominet) or .fr (delegated to AFNIC).
Most TLDs allow (for a fee) individuals and organization to get a second-level name of their choice delegated to their own DNS zone.
Each TLD is responsible for the policies governing their zone.

DNS servers use a zone configuration file to describe their RRs. The Start of Authority (SOA) record declares a new authoritative zone, by indicating its primary DNS service. Conversely, a nameserver (NS) record delegates the authority for a name suffix to another name service.
Common RR types include A and AAAA records for IPv4 and IPv6 addresses, CNAME records for aliasing, MX records for email services, etc. 
Records also include a time to live (TTL) indication, which indicates how long DNS clients may cache a given RR.
This feature is essential to DNS scalability, as users are not expected to recursively contact authoritative DNS from the root to resolve a domain name.
Instead, clients talk to a local resolver (managed by their network administrator or ISP) that will cache many of the uppers layers of the DNS hierarchy.
Then, if a client asks for the IP address of \name{www.usenix.org}, it may already have cached the NS record for \name{usenix.org} in the \name{org} zone and the SOA for \name{usenix.org}, and can directly ask the right authoritative server.

DNSSEC was proposed in the late 90s to ensure the authenticity of DNS information.
It introduces a new public key infrastructure (PKI) that follows the zone delegation structure of DNS.
Each zone can declare a key signing key (KSK), whose hash is recorded in the parent zone as a delegated signer (DS) record.
The KSK signs DNSKEY records containing the zone signing key (ZSK) effectively used to sign record sets (RRset), and these signatures are distributed in RRsig records.
To validate a RRsig, a resolver needs the DNSKEY records for the ZSK and KSK and the parent DS records for all delegations in the name. 
Finally, DNSSEC supports proof of non-existence records (NSEC3) to prevent censorship.
Although DNSSEC adoption has been slow~\cite{osterweil2008quantifying,lian2013measuring} 
, it is now supported by most TLDs and resolvers, and gaining momentum, growing from 10,000 zones in 2010 to 1M in 2020 and 10M in 2024~\cite{secspider}.

DNS-Based Authentication of Named Entities (DANE) introduces TLSA records that indicate what public key or certificate should be used to communicate with a named service~\cite{rfc6698}.
In TLS, DANE keys can either be used together with X.509 certificates for key pinning or directly for authentication~\cite{rfc7250}, replacing PKIX with the DNSSEC PKI.
DANE has gained significant adoption (750,000 zones found by SecSpider in 2024) especially for mail services. For a short time DANE was even enabled in Chrome~\cite{chromedane}.

\mysubsection{Confidential Computing}
Confidential computing leverages hardware capabilities to create TEEs that isolate its code and data from the rest of the system, including privileged components like the host operating system and the hypervisor. Because isolation is enforced implemented in hardware, a TEE can be trusted to protect the confidentiality and integrity of the code and data it processes, even from the CSP who manages the cloud hosting software stack. 
To establish trust, confidential-computing hardware provides remote attestation: TEEs can ask the hardware to sign a message together with a digest of its initial configuration (code and data) with a key only accessible to the hardware. The signature is backed by a platform certificate issued by the hardware provider. Through remote attestation a client can verify a TEE's trusted computing base (TCB) and hardware platform.

Modern CPUs from Intel, AMD, and Arm provide confidential computing capabilities. Intel Software Guard Extensions (SGX)~\cite{sgx} support the creation of in-process isolated memory regions. AMD Secure Encrypted Virtualization (SEV-SNP)~\cite{sev2020strengthening}, Intel Trusted Domain Extensions (TDX)~\cite{cheng2023intel}, and Arm Confidential Compute Architecture (CCA)~\cite{cca} provide VM-based TEEs that isolate virtual machines. Accelerators such as Nvidia GPUs~\cite{dhanuskodi2023creating} and Graphcore IPUs~\cite{vaswani2023confidential} also provide confidential computing capabilities.

Most existing TEE applications like Signal contact discovery~\cite{signalcontacts} and key recovery~\cite{signalrecovery} rely on application-level attestation checks, using a simple locally configured attestation policy such as verifying the SGX enclave signer with a fixed key. 
One challenge with this approach is that the client opens a transport session (typically using TLS) with a service that is not yet trusted, and must thus request and check attestation evidence before sending any secret data over the same session.
To stop a man-in-the-middle attack between the client and the TEE, for instance by a malicious service owner that can get a valid TLS certificate for the service,
it is also necessary to bind the attestation to the TLS handshake, a mechanism generally referred to as \emph{remotely-attested TLS} (RA-TLS) (see \S\ref{sec:related} for a survey).
RA-TLS protocols fall in two groups: either they modify TLS to bind the attestation, breaking compatibility with legacy clients, or they embed attestations in certificates,
breaking compatibility with existing CAs (and thus with legacy clients).

\begin{figure*}[t!]\centering
  \includegraphics[width=1.9\columnwidth]{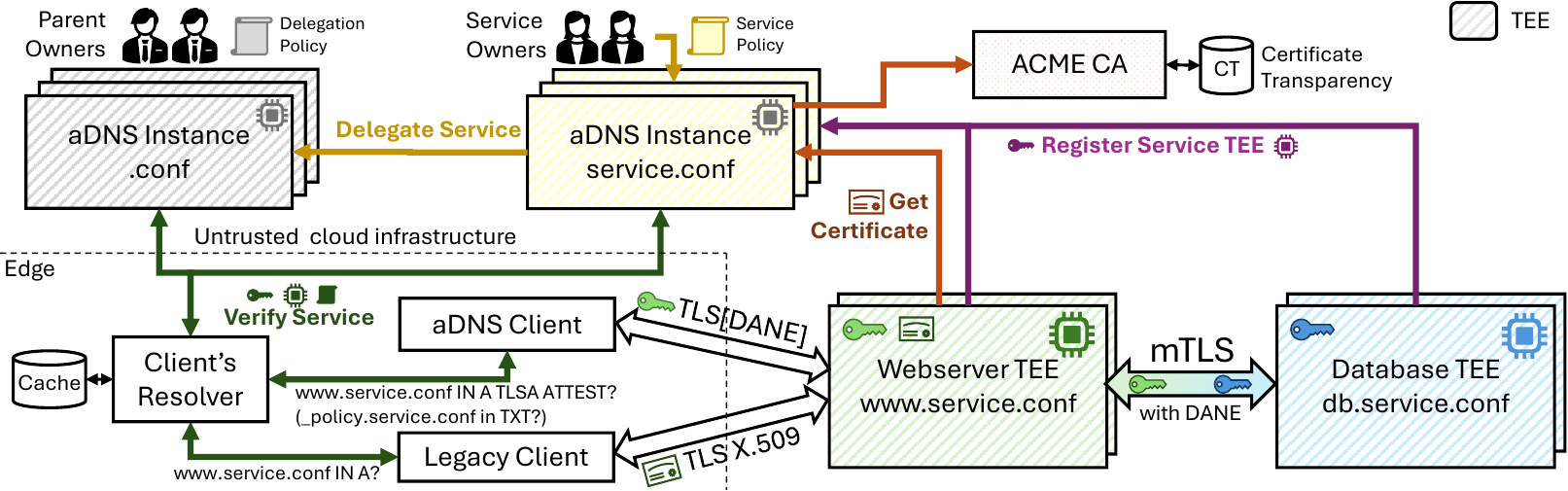}
  \vspace{-0.5em}
  \caption{Architecture of Attested DNS}
  \vspace{-0.5em}
  \label{fig:architecture}
\end{figure*}

\mysubsection{Transparency}
Transparency is a mechanism to force authorities to be consistent, auditable and accountable. Its most successful implementation is the certificate transparency system~\cite{certificate-transparency-2014} (CT) to keep certificate authorities from issuing bad certificates without getting caught. 
Transparency relies on a public verifiable append-only ledger to record information. This ledger can be routinely checked for consistency, or deeply audited in case an incident is suspected or discovered. 
Code transparency~\cite{contour18,delignat2023should} similarly uses a transparency ledger to record information about how a given binary has been built, e.g. from which source files, in which build environment, etc.

In aDNS, we are interested in the transparency of DNS records to make confidential services transparent.
Our implementation of aDNS is based on a the Confidential Consortium Framework (CCF)~\cite{russinovich2019ccf}, which operates a ledger between replicas hosted in SGX enclaves.
CCF also offers independently verifiable proofs of ledger inclusion, which can be used for holding its replicas and their operator accountable, even in case of TEE compromise~\cite{shamis2021pac}. We use the CCF ledger to make the aDNS zone configurations transparent.

\section{Architecture Overview}
\label{sec:arch}

Figure~\ref{fig:architecture} shows the high-level architecture of Attested DNS (aDNS).
At the bottom right, a sample confidential service is composed of two kinds of nodes: a front-end Web server and a database, each running in its own trusted execution environments (TEE). There are many types and implementations of TEEs (see \S\ref{sec:background}) but from an aDNS perspective we only assume that the TEEs can produce remote attestation reports.

The first core idea of aDNS is to leverage existing DNS infrastructure to distribute and cache attested information, such that enlightened clients can obtain and validate them in parallel with TLS connection establishment to services, by piggybacking on existing DNS requests to avoid incurring extra latency.
For simplicity, our presentation assumes that each confidential service is controlled by its own aDNS instance (that it, its own set of authoritative DNS servers) even though, in practice, many services can share the same zone.  
Hence, when our sample service is set up, a new aDNS instance is created for the zone \name{service.conf} (Figure~\ref{fig:architecture}, top center). 

When a TEE starts for the service, it must first register its attestation report in its zone. 
For example, a TEE running a Web server at IP address~$A$ with public key $K^+$ in a TLS certificate for domain $D={}$\name{www.service.conf} first generates an attestation report $Q$ that measures its code and configuration including $(A,K^+,D)$.
The TEE then registers $Q$ at its aDNS instance, which validates $Q$ against the registration policy on record and then creates (or updates) the DNS records for~$D$: it adds an address record $D\mapsto A$, 
an attestation record $D\mapsto Q$, and a TLSA record that pins the public key $K^+$, thereby endorsing its use to authenticate $D$.

When a client connects to \verb`https://www.service.conf`, it must at least resolve the address of the service. Modern clients also query additional DNS records, such as SVCB records to check for HTTP/3 or QUIC support, or DANE records for the TLS public key of the server. We extend aDNS-aware clients in a similar way, such that they also query for attestation records in the same DNS request.

Since the attestation records are obtained from an otherwise untrusted DNS infrastructure, they needs to be authenticated, so aDNS relies on DNSSEC to ensure its records are not tampered. But the aDNS operator could also be malicious, and may tamper with the binding between domain names and attestation reports, or sign spurious records. Therefore, the aDNS server itself must run as a confidential service protected by TEEs, and enlightened clients must also check its own attestation before trusting its records.

The second core idea is to make the TEE registration policies of confidential services discoverable and transparent. In general, if a client connects to a confidential service for the first time, it will be impossible to interpret the attestation report, as the client has no idea what code the service is meant to run, or on which TEE platform.
We introduce registration policies, which describe the conditions an attestation report should meet in order to be recorded on the DNS server. The registration policy of a service can also be fetched from DNS to be cached and inspected at the client, to help users decide if they trust a particular service or not, before the client even connects to the server. 


The third core idea is to rely on hierarchical trust in the DNS tree to deal with DNS service attestation, so that clients only need to establish trust in a small set of root aDNS instances with well-known names.
For simplicity, our presentation assumes a single root aDNS instance that operates the \verb`.conf` top-level domain (TLD), but in practice many attested zones may co-exist in the name hierarchy. 
Hence, we have the root zone (managed by ICANN with the root DNSSEC keys), an attested TLD zone \name{conf} (top-left in the figure, controlled by an aDNS instance with its own attested KSK), and a attested service zone \name{service.conf} (top-center, controlled by another aDNS instance with its own attested KSK).
Any time a sub-zone is delegated to a new aDNS instance (e.g. \verb`service.conf`), it is the responsibility of the parent aDNS to verify the attestation of the new instance. Therefore, aDNS instances are configured with a {\em delegation policy} that describes what constitutes valid implementations of aDNS and valid TEE platform to run them. Delegation policies are applied transitively through delegation, so they can only become more restrictive at each delegation step. For instance, the TLD instance may allow implementations of aDNS running on Intel, AMD or Arm CPUs but the delegation policy for \verb`service.conf` may only allow Intel or AMD CPUs.

Finally, a major benefit of aDNS is the ability for a TEE to take a dependency on other services (inside or outside its DNS zone), for instance a database at \name{db.service.conf}, without having to configure their credentials.
%
Today, services that want to use mutually-authenticated TLS require a control plane (e.g. a service mesh~\cite{8705911} like as OpenShift and a proxy such as Envoy~\cite{envoy}) to provision client certificates and deal with authentication and authorization. With aDNS, both endpoints can use their DANE raw public keys, so that trust is established on the basis of the policy enforced for each service's name, without having to deal with any certificate or attestations in application code. 

\section{Security Goals and Threat Model}
\label{sec:threat-model}

\noindent{\bf Core security goal.}
A confidential service comprises a DNS zone (e.g. \name{service.conf}) and a registration policy that specifies the rules that a TEE must satisfy to run the service under a given name in this zone (e.g. \name{www.service.conf}).
The main goal of aDNS is to ensure that, if a client successfully establishes a TLS session to a server with such a DNS name (e.g., a Web server), then this server must run in a TEE with an attestation that meets the associated registration policy.

The registration policy and service code are authored by a service owner who is responsible for authoring them in line with users' expectations (e.g., to ensure privacy of data sent to the service).
The service owner is trusted, but their actions are auditable through aDNS's transparency log.
We assume the hardware-based TEE design, implementation, and manufacture are correct, and trust the PKI for endorsing device certificates and TCB updates (e.g. firmware).
The service is deployed and operated by an untrusted service operator that may compromise any components outside TEEs (such as host operating systems, hypervisors, devices, networks and non-confidential services).

\begin{table}\scriptsize
  \addtolength{\tabcolsep}{-0.1em}
  \begin{tabular}{|llllll|}
  \hline
  \multicolumn{1}{|c}{\textbf{\begin{tabular}[c]{@{}c@{}}Client \\ Type\end{tabular}}} &
    \multicolumn{1}{c}{\textbf{Resolver}} &
    \multicolumn{1}{c}{\textbf{\begin{tabular}[c]{@{}c@{}}Web \\ PKI CA\end{tabular}}} &
    \multicolumn{1}{c}{\textbf{\begin{tabular}[c]{@{}c@{}}DNSSEC \\ Root\end{tabular}}} &
    \multicolumn{1}{c}{\textbf{\begin{tabular}[c]{@{}c@{}}Service's \\ aDNS\end{tabular}}} &
    \multicolumn{1}{c|}{\textbf{\begin{tabular}[c]{@{}c@{}}Parent's \\ aDNS\end{tabular}}}
    \\ \hline\hline
  \textbf{aDNS Client} &
    Untrusted &
    Untrusted &
    Untrusted & 
    Untrusted &
    Trusted \\ \hline
  \textbf{\begin{tabular}[c]{@{}l@{}}aDNS Client \\ + discovery\end{tabular}} &
    Untrusted &
    Untrusted &
    Untrusted & 
    \begin{tabular}[c]{@{}l@{}}Trusted \\ Transparent\end{tabular} &
    Trusted \\ \hline
  \textbf{DANE} &
    Untrusted &
    Untrusted &
    Trusted &
    Trusted &
    Trusted \\ \hline
  \textbf{X.509} &
    Trusted &
    \begin{tabular}[c]{@{}l@{}}Trusted \\ Transparent\end{tabular} &
    - &
    Trusted &
    Trusted \\ \hline
  \end{tabular}
  \vspace{-1em}
  \caption{Trust assumptions for different kinds of clients}
  \vspace{-1em}
  \label{tab:trust-assumptions}
\end{table}

aDNS is designed to provide this core guarantee to a broad range of clients with different capabilities.
Table~\ref{tab:trust-assumptions} summarizes the trust assumptions required for each type of (correctly implemented) client for our core security to hold.


A fully-enlightened aDNS client (implemented, e.g., as a local resolver or a browser extension) fetches attestation from DNS and verifies it against a trusted local attestation policy.
Although these clients rely on aDNS as an untrusted cache for availability and freshness, their security doesn't depend on the service owner or operator. In other words, they are as secure as today's service-specific clients that fetch and verify attestation at the application layer or use RA-TLS.

Next, we consider an aDNS client with the same capabilities but no {\em a priori} knowledge of the service it connects to, which we refer to as "discovery" because the client must discover the registration policy and either verify its authenticity (using, e.g., signed policies) or trust that the parent aDNS instance only accepts policies from authorized service owners.
In the latter case, the client trusts both the service owner and aDNS, but it can rely on transparency and other mechanisms to mitigate that trust. 
The client may present the registration policy to the user to review, and retain it for future connections and audits. 
The client may also establish trust in aDNS instances by fetching and verifying their attestation, and by pinning their KSKs; these options are discussed in \S\ref{sec:clientproto}.

Finally, we consider two kinds of aDNS-unaware clients that 
also benefit from aDNS's attestation policy enforcement.
Clients that support DANE for TLS authentication will check the peer's key is registered on aDNS, at the cost of additional trust in DNSSEC for the authenticity of TLSA records, and in aDNS instances for verifying attestation reports of TEE instances before publishing TLSA records.
These clients also trust the parent aDNS for correct delegation to the service aDNS. However, no trust is required in Web PKI CAs since these clients authenticate services based on TLSA records instead of certificates.

Clients that solely rely on X.509 certificates issued by Web PKI CAs must additionally trust that these CAs issue certificates only after authenticating CAA records using DNSSEC and challenging TXT records (\S\ref{sec:acme}).
as prescribed by the CA/Browser Forum Baseline Requirements~\cite{cab}.
While rogue certificate issuance cannot be prevented, CAs can be held accountable using Certificate Transparency and logging certificate issuance in aDNS.

\mysubsection{Additional security properties}
aDNS provides two additional properties to some clients.

{\em Service transparency}: all clients that use at least DNSSEC are guaranteed they are served the same records (including the service policy) as every other clients as long as aDNS is not compromised. Each aDNS instance also maintains a tamper-proof ledger recording all changes to the records in the zone---this allows an audit of the full historical TCB of both the service and its aDNS instance, and also provides residual security in case of TEE compromise.   

{\em Freshness and linearity}: aDNS clients (with or without discovery) are guaranteed to connect to up-to-date registered TEEs with fresh attestation evidence assuming clocks are synchronized and TEEs have access to a trusted time source.

\section{Protocols for Confidential Services}
\label{sec:protocols}

\subsection{Registering a TEE for a Service}\label{sec:registration}

We first describe the protocol to provision a new TEE to run as part of a confidential service, and more generally discuss the attestation-based authorization policies enforced by aDNS. The creation and registration of the service itself in an attested DNS zone are explained in \S\ref{sec:delegation}. 

\medskip\noindent{\bf Attestation reports} presented by service TEEs are of the form 
$Q[\mathit{platform},\mathit{code},\mathit{config}; \vec K^,\mathit{time}]$
where   
\begin{itemize}
\item $\mathit{platform}$ includes platform-specific information, such as its hardware identity and firmware patch level; we keep the details and the actual hardware signature abstract; 
\item $\mathit{code}$ is a digest of the code running in the TEE: based on the attestation, the verifier can infer that any security invariant programmatically enforced by that code holds for the TEE that issues the report;  
\item $\mathit{config}$ is the configuration of this TEE as part of an instance of the service; it notably includes the domain name and a unique identifier for the service instance, e.g. $D = {}$\name{service.}\apex{} and \verb$v0.1$; 
a unique prefix for this TEE as part of this service, e.g. \name{node42.front-end}; 
and the list of origins to be served by this TEE, including their prefixes, IP addresses, protocols,  ports, and key indexes, e.g., \name{www} : \verb${https, 443, 1}$.  
\item $\vec K^+$ is a list of public keys presented by the TEE.
Each key is labelled with a signature algorithm and a usage---either DANE or X.509. The first key is always the DANE key used to uniquely identify this TEE;
\item $\mathit{time}$ is the attestation time presented by the TEE.  
\end{itemize}
Intuitively, the attestation report transparently binds the name of the TEE within the service, e.g. \name{node42.front-end.$D$}, to its hardware platform, its software configuration, and its fresh identity key, and its similarly binds the origins it serves to fresh authentication keys.  

\medskip\noindent{\bf Registration}\label{sec:register}
enables a TEE provisioned by an untrusted service operator to join a confidential service, as follows. 

\begin{enumerate}
  \setlength\itemsep{0em}
    \item The service operator creates a new TEE for its choice of hardware platform, $\mathit{code}$ and $\mathit{config}$.     
    
    \item\label{step:start} The TEE starts executing this code, which presumably\footnote{Otherwise attestation verification fails at step 3.}
    \begin{enumerate}
    \item samples its fresh private keys, records its host time, obtains a hardware-attested report of the form given above, and gathers any supporting evidence, such as platform certificates. 
    
    \item retrieves and verifies the server keys and attestations of the aDNS instance of the service, as detailed in  \S\ref{sec:clientproto}, based on the DNS name in $\mathit{config}$.

    \item \label{step:raw} opens a mutually-authenticated TLS connection with aDNS, presenting its DANE identity key as client key and verifying that aDNS presents one of its retrieved attested key as server key. 
        
    \item posts a REGISTER request with its attestation report and supporting evidence as payload.  
    \end{enumerate}

    \item aDNS accepts the connection, 
    verifies the attestation report, 
    checks that its claimed service name is in the service zone it controls, 
    that it passes the registration policy on record for this name, 
    that its time is reasonably recent, 
    and that its primary DANE public key matches the raw key presented in step \ref{step:raw}, 
      
    If all these checks succeed, aDNS then installs new ATTEST, TLSA, and A/AAAA records for each of the origins declared in the attested configuration. In particular, it installs a TLSA record for DANE authentication using its primary key at its unique name. 
    (Any clashes with existing records cause the registration to fail.)
 
    \item Once aDNS acknowledges its registration, the TEE starts running on behalf of the service. For any subsequent connection with aDNS, it authenticates as a DANE client using the TLSA record for its identity key.  
    
\end{enumerate}
The TEE may then request certificates for some of its origins (see \S\ref{sec:acme}).  
The TEE may also update its registration with a fresh attestation report (e.g. to periodically rotate its keys) resuming from steps \ref{step:start}, subject to the service registration-update policy. The main difference is that old records are removed before installing new records for its updated origins. 

\cedric{TODO: be more precise about the records we add. Either here, on by example for L7.}

\cedric{TODO: explain what policies should enforce; I am concerned about minimal policy checks to defend the zone, e.g. DELEGATE/REGISTER label clashes.}

\smallskip\noindent{\bf Security\;} 
For any successfully registered TEE, aDNS must have verified and logged a corresponding attestation report that passed the service policy at the time of registration. 
Since the primary DANE key must be freshly sampled, the registration report uniquely identifies honest TEE instances.
Further, since aDNS verifies possession of the DANE private key in the registration protocol (by requiring mutually-authenticated TLS), attackers cannot re-play a registration request unless the TEE is compromised or its code is malicious. 
Service owners may want to limit how long a registration is valid for and require TEEs to re-attest regularly.
We could require the report to include a random challenge by aDNS to ensure freshness of the report; however, since our client freshness guarantees are time-based, we instead rely on attested time. Most TEE implementations do not provide a secure time primitive, but we can rely on signed timestamps from an external trusted time source to limit how long stale reports can be used.

\subsection{Obtaining Certificates for Registered TEEs}
\label{sec:acme}
    
\cedric{From the RFC, it is unclear whethere there is a distinct challenge for each FQDN.}

Some TEEs need Web PKI certificates in addition to DANE to support legacy clients, so aDNS also supports 
a protocol to obtain certificates from any ACME-compliant CA. 
This protocol may run at any time after TEE registration; 
its goal is to ensure that certificates for names in the service zone endorse only keys of registered TEE instances, to prevent e.g. a malicious service owner (who owns the service's domain) to attack legacy X.509 clients.

\mysubsection{ACME and CAA (background)}
Automatic Certificate Management Environment (ACME, RFC 8555~\cite{rfc8555}) is a standard
that enables certificate authorities (CAs) to verify that a certificate request legitimately represents the domain names to be endorsed in the resulting certificate without human intervention. 
It is supported notably by Let's Encrypt and thus used to issue the majority of Web certificates, endorsing more than 400M domain names to date. 
%
ACME can be configured to verify ownership of domain names by requesting inclusion of a specific TXT record bound to a fresh random challenge for each of the requested domain names. 

Certification Authority Authorization Resource Records (CAA, RFC 8659 and 8657) is a complementary standard 
that defines records to indicate which CAs and verification methods should be used for certificate issuance in a zone; it is considered best practice by most root CAs. 
The combination of CAA and ACME allows us to limit the TCB of certificate issuance for confidential services to their authoritative aDNS instance and their designated CA.  

\begin{figure}[!t]
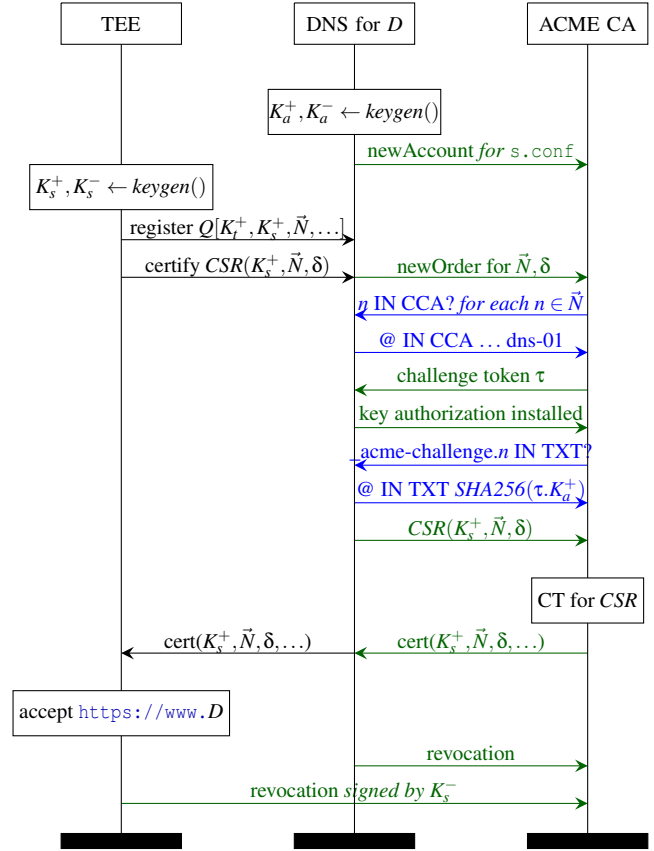

  \begin{msc}{} \footnotesize
  \drawframe{no}
  \setmsckeyword{}
  \setlength{\labeldist}{0.2ex}
  \setlength{\topheaddist}{0cm}
  \setlength{\instdist}{3.1cm}
  \setlength{\envinstdist}{0pt}
  \declinst{s}{}{TEE}
  \declinst{d}{}{DNS~for~$D$}
  \declinst{a}{}{ACME CA 
  }
  \action*{$K^+_a, K^-_a \leftarrow \mathit{keygen}()$}{d}
  \nextlevel 
  \nextlevel
  \action*{
    $K^+_s, K^-_s \leftarrow \mathit{keygen}()$ }{s}
  \color{darkgreen}\mess{newAccount \emph{for} \name{s.conf}}{d}{a}
  \color{black}
  \nextlevel[2]
  \mess{register $Q[K^+_t, K^+_s, \vec N, \dots]$}{s}{d}
  \nextlevel
  \mess{certify $CSR(K^+_s, \vec N, \delta )$}{s}{d}
  \color{darkgreen}\mess{ newOrder for $\vec N, \delta$}{d}{a}
  \nextlevel 
  \color{blue}\mess{$n$ IN CCA? \emph{for each} $n \in \vec N$}{a}{d}
  \nextlevel
  \mess{@ IN CCA $\dots$ dns-01}{d}{a}
  \nextlevel
  \color{darkgreen}\mess{challenge token $\tau$}{a}{d}
  \nextlevel
  \mess{key authorization installed}{d}{a}
  \nextlevel
  \color{blue}\mess{\_acme-challenge.$n$ IN TXT? 
  }{a}{d} 
  \nextlevel
  \mess{@ IN TXT $\mathit{SHA256}(\tau.K^+_a)$}{d}{a}
  \nextlevel 
  \color{darkgreen} \mess{$CSR(K^+_s, \vec N, \delta)$}{d}{a}
  \nextlevel
  \color{black}
  \action*{~CT for $CSR$~}{a}
  \nextlevel
  \nextlevel
                   \mess{cert($K^+_s, \vec N, \delta, \dots$)}{d}{s} 
  \color{darkgreen}\mess{cert($K^+_s, \vec N, \delta, \dots$)}{a}{d}
  \nextlevel
  \color{black}
  \action*{~accept \url{https://www.}$D$~}{s}
  \nextlevel
  \nextlevel
  \color{darkgreen}\mess{revocation}{d}{a}
  \nextlevel
  \mess{revocation \emph{signed by}{} $K^-_s$}{s}{a}
  \end{msc}
  \vspace{-6mm}
  \caption{\em Service registration and certification, including {\color{darkgreen}ACME} (green) and {\color{blue}DNS} (blue) messages. 
  $K^+_a$ is the account key for the service implicitly used to authenticate ACME requests;
  for simplicity, we also omit the identity key of the TEE used to authenticate its register and certify commands;  
  $K^+_s$ and $\vec N$ are the Web public key and HTTPS origins included in the TEE attestation 
  and authorized by its service registration for which the TEE requests a certificate;
  $\delta$ abbreviates the requested validity interval (the $\mathit{notBefore}$ and $\mathit{notAfter}$ fields in the certificate) subject to both service and CA policies.} 
  \label{fig:acme}
  \vspace{-4mm}
  \end{figure}

\mysubsection{aDNS-based ACME} 
ACME is based on a series of JSON-encoded HTTPS requests from a client to the CA.
We omit some details, such as error handling and request state polling.
ACME clients sign all their request with an account key, recorded by the CA when it creates the account. 

For our purpose, the protocol (given in Figure~\ref{fig:acme}) involves a requesting TEE, its aDNS service acting as ACME client, and the CA recorded in the service configuration, such as \name{letsencrypt.org}, acting as ACME server. 
As part of its initialization, the authoritative aDNS for the service generates a public-private keypair ($K^+_a, K^-_a$) and creates an account for all certificate requests in the service zone---the private account key~$K^-_a$ never leaves aDNS TEEs. 
aDNS also installs CAA records to prevent any standard-compliant CA from issuing certificates except for those it will explicitly request.    
For example, this can be achieved by adding a record 
\[ \small D \mapsto \begin{array}[t]{@{}l}\verb$CAA 0 issue$ \\ \verb$"letsencrypt.org;validationmethods=dns-01"$\end{array}\] 
stating that only \name{letsencrypt.org} can issue a certificate for any name that ends with $D$ 
and that it must validate ownership of that name using a DNS challenge. 

Assume a TEE registered with a single X.509 keypair ($K^+_s,K^-_s$) configured to serve both \name{www.}$D$ and $D$ under a common certificate for $K^+_s$ (following common Web practice).
%
To request the certificate, the TEE prepares and signs a PKCS\#10 Certification Signing Request (CSR) for one of its attested X.509 keys and some of the origins recorded in its currently-registered attestation record.
The CSR includes the public key, the requested names $\vec N$, and the notBefore and notAfter fields $\delta$ to be included in the resulting certificate; 
it is signed with the corresponding private key ($K^-_s$) as proof of its possession. 
The TEE then delegates certificate issuance to its aDNS by sending a certify command for this CSR.

aDNS verifies the CSR matches the template for one of the X.509 keys in the 
current attestation record for the TEE and that every name $n \in \vec N$ ends with the service name. 
It then logs the CSR and sends an ACME newOrder request for $\vec N$ and $\delta$ signed by the confidential service account key.   

The CA verifies the request is legitimate: it first sends DNS requests to collect any relevant CAA records, starting with each full name $n \in \vec N$ and removing one prefix label at a time, until it gets a CAA record or it reaches the root. 
This crucially relies on DNSSEC to authenticate CAA records or their absence.
Accordingly, the CA generates a fresh dns-01 challenge 
and sends it back to aDNS.

Following the standard \verb$dns-01$ verification method, aDNS proves that it controls every requested name $n$ by computing an `authorization key' $k$ as the hash of the challenge concatenated with the hash of its public account key and by installing a `'\verb$TXT 300 $$k$'' record at every name \name{\_acme-challenge.}$n$.
It then signals the CA that it is ready to meet its self-assigned challenges.
(As an invariant, to prevent interference with other sub-protocols, aDNS refuses to add TXT records with the reserved label \name{\_acme-challenge} for any other purpose.) 

The CA queries each of these records, authenticates them via DNSSEC, and verifies that their payloads match the authorization key it expects for this order.
It is now ready to receive the original self-signed CSR prepared by the TEE, verify its signature, match its contents against the pending order, 
and trigger the production of a certificate that endorses this contents, in two stages:  
(1) the CA publicly registers the CSR to ensure certificate transparency; 
(2) the CA signs the new certificate using its own private signing key, and returns this certificate to our aDNS client. 
aDNS verifies that the certificate matches the CSR and complies with certificate transparency.  
It logs this certificate, installs the corresponding TLSA records, and returns the certificate to the requesting TEE to present to legacy clients for the names in $\vec N$.
%

%

\mysubsection{Security} 
The main property of interest is that a client accepts a certificate as valid for any origin 
with the name of a confidential service only 
if 
(1) the key has been attested by a TEE registered for the service; 
(2) the corresponding PKIX origin has been authorized by the service policy; 
(3) the leaf certificate has been endorsed by the CA recorded in the service configuration; and 
(4) the aDNS for the service has logged this certificate after successfully completing the issuance protocol above.
The protocol crucially depends on the security of ACME and its use of DNSSEC by the CA to authenticate both CAA and challenge TXT records. ACME CAs are encouraged to implement additional mitigations to protect against an attacker corrupting a DNSSEC key or downgrading DNSSEC (if it doesn't chain to the root keys), such as performing DNS resolution from multiple network paths to ensure they are consistent. Pinning aDNS KSKs can also protect clients from the DNSSEC hierarchy.

If a CA that doesn't implement ACME or CAA properly issues a certificate to a name in a confidential service zone (for instance, using a weak validation method such as sending a challenge to the email address of the service owner), then the service owner may attack legacy X.509 clients. A Certificate Transparency monitor can detect the issuance of such certificates to hold the service owner (or bad CA) accountable.

\subsection{Connecting to a Confidential Service}
\label{sec:clientproto}

\begin{figure}[t]
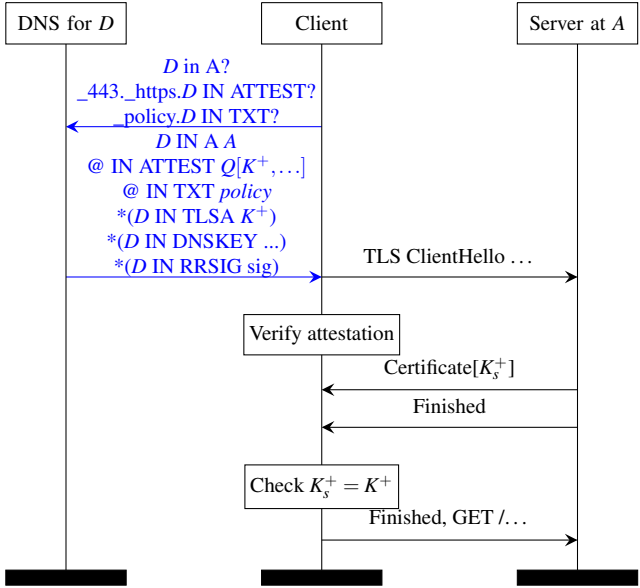

  \begin{msc}{} \footnotesize
  \drawframe{no}
  \setmsckeyword{}
  \setlength{\topheaddist}{0cm}
  \setlength{\instdist}{3.4cm}
  \setlength{\envinstdist}{0pt}
  \setlength{\actionheight}{.6\actionheight}
  \declinst{d}{}{DNS~for~$D$}
  \declinst{c}{}{Client}
  \declinst{s}{}{Server~at~$A$}
  \nextlevel 
  \color{blue}
  \mess{
    \begin{tabular}{c}
      $D$ in A? \\
      \_443.\_https.$D$ IN ATTEST? \\
      \_policy.$D$ IN TXT? \\
      \\
    \end{tabular}}{c}{d}
  \nextlevel
  \nextlevel
  \nextlevel
  \nextlevel
  \mess{
    \begin{tabular}{c}  
      $D$ IN A $A$ \\
      @ IN ATTEST $Q[K^+,\dots]$ \\
    @ IN TXT \textit{policy} \\
    {}*($D$ IN TLSA $K^+$) \\
    {}*($D$ IN DNSKEY ...) \\
    {}*($D$ IN RRSIG sig) \\
    \\
    \\
    \\
    \\
    \end{tabular}}{d}{c}
  \color{black}
  \mess{TLS ClientHello \dots}{c}{s}
  \nextlevel
  \action*{Verify attestation}{c}
  \nextlevel
  \nextlevel
  \mess{Certificate[$K^+_s$]}{s}{c}
  \nextlevel
  \mess{Finished}{s}{c}
  \nextlevel
  \action*{Check $K^+_s = K^+$}{c}
  \nextlevel
  \nextlevel
  \mess{Finished, GET /\dots}{c}{s}
  \end{msc}
  \caption{\em Enlightened aDNS client connecting to an HTTPS service endpoint at $D$ with unknown address and policy. All DNS messages are in blue.}
  \label{fig:adnscli}
  \vspace{-4mm}
  \end{figure}

Assume a client attempts to connect to a confidential service at \name{https://}$D$. 
The full process for an aDNS-aware client, shown in Figure~\ref{fig:adnscli}, relies on standard protocols. It consists of preparatory DNS queries to retrieve and validate the attested key of the server, followed by a standard TLS handshake authenticated with this key. 
  
\mysubsection{aDNS client}
The attestation report(s) for \name{https://D} 
(see~\S\ref{sec:registration} for details)
are available as {\tt ATTEST} record at {\tt \_443.\_https.}$D$.  
As with {\tt SVCB} service binding records (RFC 9460~\cite{rfc9460}), the protocol and port are encoded in a prefix to enable multiple services and instances at a given name.\footnote{We could also distribute attestations using key-value pairs in SVCB records for additional compatibility with legacy resolvers.}

If an aDNS client is connecting to a service for the first time and doesn't have a local service policy (discovery), it may also query for the service policy (available as a \name{TXT} record at \name{\_policy.D}), and recursively, the attestation evidence of the service's aDNS instance. The client may ask the user to authorize the policy directly, or trust the decision of the parent's aDNS to authorize the delegation.

The query for {\tt ATTEST} and {\tt TXT} records may be bundled with basic DNS address resolution ({\tt A} and {\tt AAAA} records) in a single packet.
Reports and addresses are indexed the same way so the client knows which instance to expect when connecting to any address.
Responses are recursively cached at every resolver (including at the client) and may return additional relevant records (with asterisks in the figure). 

The client can initiate the connection to the service in parallel with the verification of the service's attestation report. However, clients must be careful to send 0-RTT data only after the attestation has been verified.
The client must also check that TLS server key matches one of the keys in the attestation report. 
%
The client is expected to cache the result of attestation verification (and optionally, the authorized policy) for the minimum duration of either the TTL of the ATTEST record, or the maximum configured duration of the client.

\mysubsection{Legacy client}
Along with the DNS address resolution response, if the resolver supports DNSSEC, it will receive additional records to authenticate the response: this includes the RRSIG record of the A (and optionally TLSA) RRset, the DNSKEY records of the service zone, and recursively, the chain of DS records up to either a pinned trusted aDNS root instance, or a global DNSSEC root key.
If the client supports DANE, it should query the \name{TLSA} record and check one of its values matches the TLS raw public key or certificate presented during the TLS handshake with the service.
Otherwise, the client will validate the certificate using its set of trusted root CAs and ensure it includes the name of the service.

\cedric{Does this also apply to clients connecting to aDNS instances? They can get records (including attestation reports) via DNS queries, and additionally rely on DoH/DoT, whereas other commands are REST-based. For instance, it may be worth discussing what an enlightened aDNS resolver would do.}

\mysubsection{Security} We consider each of the clients in Table~\ref{tab:trust-assumptions} in turn, first 
addressing how they meet our main security goal, then discussing additional freshness and transparency guarantees.

An {\bf aDNS client with a local service policy} is as secure a client using any variant of attested TLS. 
The main difference between the two is that the server attestation is received in a signed but a priori untrusted DNS record, instead of an priori untrusted TLS connection, but it does not matter as long as this attestation complies with the service policy and authenticates the server TLS key. 
%
%
%
%
If the client and aDNS clocks are (roughly) synchronized and the client's resolver checks DNSSEC is not expired, then the aDNS client also knows the instance it connects to has a fresh attestation for an up-to-date policy (up to TTL on its attestation record). 
%

An {\bf aDNS client discovering the service policy} either 
knows how to locally verify the (a priori untrusted) service policy retrieved from DNS---in which case it is equivalent to the case above---or it must trust that aDNS correctly authorized and authenticated this policy for this service name, with different subcases depending on zone delegation. 

If a client already trusts an aDNS instance (and in particular its delegation policy) for a prefix of the service name, 
then, by recursion over each delegation step to the aDNS instance for the service, it can also trust this service policy. 
As with any other confidential service, trust in an aDNS instance can be achieved 
by fetching and verifying an attestation for this instance against a locally-trusted aDNS-verification policy, and 
then pinning its attested KSK for DNSSEC validation of any record in its zone of authority. 
Otherwise, if the client also discovers the whole attested zone for the service, then it must also trust the DNSSEC PKI to authenticate the KSK of the aDNS instance at the apex of this zone. 

%

%
A {\bf legacy DANE client} delegates attestation checking to aDNS, but as long as aDNS is trusted (as discussed above), 
this is equivalent for our main goal to an aDNS client that discovers then locally verifies the policy, 
and it is slightly weaker for transparency inasmuch as the DANE client cannot log or review the policy applied. 

A {\bf legacy X.509 client} additionally depends on the security of the certificate issuance protocol, discussed in \S\ref{sec:acme} above.  
In other words, if trusted CA issue certificates for a confidential service only after CAA and ACME DNS-based name validation,  then these clients get the same security as DANE clients, minus the freshness guarantees of DNSSEC.

In practice, the trust assumptions for legacy clients are also mitigated when some aDNS clients verify attestation reports of aDNS instances, check aDNS and CT logs, and report inconsistencies. For example, by verifying attested evidence for aDNS, aDNS clients can ensure it serves the same records to all clients. In particular, any use of registration policies that do not meet end-user expectations or permit code with known vulnerabilities can be detected by aDNS clients or auditor.

\section{aDNS Services and Zone Delegation}\label{sec:delegation}

\noindent{\bf Creating an aDNS Instance.} 
The protocol for starting an instance of aDNS in the DNS zone hierarchy 
is a refinement of the protocol for adding a TEE to a confidential service (\S\ref{sec:registration}). 
The operator creates a TEE that runs the initial primary DNS server, for its choice of code and configuration. 
(This may involve DNS requests to the intended parent aDNS instance to make sure the child and parent configurations
are aligned, and prevent later failures.)

As it starts, the new aDNS TEE samples fresh keypairs and produces an attestation of the form 
$Q(\mathit{platform},\mathit{code},$ $\mathit{config}; \mathit{KSK}^+, K^+_{ns0}, \mathit{time})$
where $\mathit{config}$ sets 
the intended full name of the zone (e.g. $D = {}$\name{service.conf}),
the new zone's policies, and 
the parent DNS presumably willing to delegate this zone.
The configuration also indicates whether the parent runs aDNS or just DNSSEC, 
in which case the new zone is the apex of a island of attested names in the DNSSEC PKI.

The attestation includes two fresh public keys: 
its initial DNSSEC key-signing key $KSK^+$ and a TLS authentication key $K^+_{\mathit{ns0}}$ 
used to serve both DNS queries over HTTPS and TLS (for reading resource records) 
and aDNS commands over HTTPS (for modifying them).

We name all authoritative DNS servers using reserved labels of the form \name{ns}$n$ for $n \geq 0$  prefixed to the zone name. As outlined in the implementation (\S\ref{sec:impl}), the startup process may also involve 
mutually-attested communications with auxiliary TEEs that run backup DNS servers for the new zone.

The policies in the configuration are used to guard every command that may update the resource records in the zone or its authoritative DNS servers.
This includes, for instance, commands for service TEE registration, sub-zone delegation, certificate issuance, code update, and policy update. 
In this presentation, we make the following simplifying assumption:   
aDNS instances are either `leaf' zones for TEE registration, 
or `intermediate' zones 
that only support delegation.
While not essential, this prevents conflicts between registration and delegations under the same name. 

Unless it is the apex of an attested island, the new aDNS instance sends a delegation request to its parent, 
as detailed below. 
Once accepted, it gets its Web key certified by the CA indicated in its configuration, as detailed in \S\ref{sec:acme}, acting both as requesting TEE and ACME client. 

\mysubsection{Delegating a Zone to a Child aDNS Instance}
We now explain how the parent aDNS handles delegation requests.
The request payload mostly consists of the child's attestation report, 
together with additional DNS information provided by the primary
such as glue records containing the IP addresses of its DNS servers to be advertised by the parent. 
It is sent over mutually-authenticated TLS using DANE keys.
Before opening the connection (see \S\ref{sec:clientproto}), the child requests and 
verifies the TLSA and ATTEST records to confirm the authenticity of the DANE key. 
In contrast, since the child DNS service is not yet part of the DNSSEC hierarchy, 
the parent provisionally accepts connections authenticated by \emph{any} client key and then
verifies that this key matches the attested DANE key in the report 
at the application level. 

Upon receiving a new delegation request, the parent first verifies 
that it passes its own delegation policy. This involves 
verifying the child attestation report,
checking that its code and configuration are correct for a new aDNS instance, 
checking that the name requested by the child is a subzone of the parent zone
and that it has not yet been delegated. 
It also verifies that the child's delegation policy is compatible with its own (and ancestors') policies (see below).
%
Assuming all these checks succeed, the parent updates its records to implement the delegation:  
following DNS and DNSSEC specifications, it 
creates a NS record for the child, a DS record for the child's attested key ($KSK^+$ in $Q$)
at the child name, and glue records. 
It also creates an ATTEST record that logs the child attestation, making it transparent and auditable. 
The child waits for the confirmation that it delegation request has been accepted, then 
starts serving DNS requests and aDNS commands. 

Later on, the child may send delegation-update requests to refresh its attestation report, its DNSSEC key, or just update its glue records. 
The parent similarly verifies then executes these requests; 
the main difference is that the child authenticates with a DANE key on record in the attestation 
currently held by the parent. (This prevents conflicts with fresh delegations.) 

\smallskip\noindent{\bf Policies for Delegation and Code Updates}
We finally present a policy design that strikes a reasonable balance between the privileges of child and parent aDNS instances. 

From a classic DNS viewpoint, the operator of a parent zone retains strong discretionary powers over its delegated sub-zones. In particular, it can modify or delete their delegation records, potentially endangering their security.
With aDNS, in contrast, the owner of the parent zone only fixes its initial policies (including its ability to update them), which are then automatically enforced. This defines and limits the scope of discretionary updates, and ensures that these policies and their consequences will be transparently logged for audit. 

To enforce meaningful security invariants across large sub-delegated zones,  
e.g. to enforce that all its aDNS TEEs meet minimal platform and code requirements,  
it is important to limit the scope of the policies that will be recursively enforced by child aDNS instances.
To this end, we propose that the child delegation policy (including the policy to verify children aDNS attestations) be at least as strict as their parents'. 

Our implementation enforces it by including in the child aDNS configuration the policies of all its parents up to the apex of the attested island,
by checking this inclusion as part of the verification performed by the parent during delegation, and by independently evaluating \emph{all} the delegation
policies in the aDNS configuration before accepting delegation requests.
In contrast, delegated instance have full control over their TEE registration policies for any other service besides aDNS. 

\section{Implementation}
\label{sec:impl}

\subsection{Attestation library}
To support a broad variety of TEEs, we need a library that can verify different types of reports and is highly portable to run both in
the aDNS server and in many clients.
We could delegate attestation verification to services provided by CSPs and hardware manufacturers, but this would 
add a highly trusted entity to our TCB.
Instead, we implement Ravl, a new remote attestation validation library that only depends on OpenSSL.
Ravl introduces a universal attestation format, consisting of a TEE type, a main report data, and a collection of tagged collaterals for certificates,
CRLs and OCSP proofs, RIM bundles or custom collaterals such as Intel TCBInfo.
Ravl is also capable of fetching outdated or missing collaterals, and offers historical validation (validity at a given timestamp).

\subsection{aDNS Server}
We implement an aDNS server using the Confidential Consortium Framework~\cite{russinovich2019ccf} (CCF), which runs in Intel SGX enclaves.
We choose SGX as it remains the smallest TCB option among TEE implementations offered by mainstream cloud providers.
CCF simplifies the implementation of aDNS thanks to several of its built-in features. Some of these features are mainly for convenience,
such as the fact that CCF applications only need to implement application endpoints and the framework takes cares of mundane tasks such as
managing host to enclave communication, TLS sessions, or HTTP requests. Other features solve difficult issues for us.

First, CCF provides operational reliability. 
Most DNS services require at least 2 instances for availability, but maintaining 
consistency between instances can be hard. Traditionally, one of the instance is the primary while backups use the zone transfer features of
DNS to copy the primary's configuration. Authorizing and managing backups is even more challenging in the confidential computing setting
where enclaves may be forked and kept alive by an attacker to keep stale configurations live. 
CCF implements a variant of the Raft~\cite{ongaro2014search} consensus to synchronize the leader with backup nodes.
It manages mutual attestation between its replicas and takes care of reconfiguration automatically if the primary fails.

Second, CCF produces a tamper-proof ledger of transactions. The ledger is used internally by CCF to ensure the persistence and recoverability
of the state of the application (especially if all replicas crash, a scenario referred to as disaster recovery). CCF  maintains both the integrity
(with a Merkle tree) and confidentiality (by encrypting entries with keys only accessible to TEEs) of its ledger. We take advantage of its ledger
to provide auditability and transparency of confidential services served through aDNS. Every change to the zone configuration (e.g. registering
a TEE) is recorded as a transaction in the ledger. Therefore, if an aDNS server publishes its ledger, an auditor can check the history of
all TEEs that ever participated in a service. Another benefit of the ledger is that it enables {\em receipts}, consisting of a Merkle path 
and a signed root hash to prove that a given transaction was recorded in the ledger at a given state. We use receipts to verify a fresh aDNS instance
is properly configured for delegation.

Third, CCF has an explicit concept of the {\em governance consortium} of a service, that is, the group of people that collectively represents the trusted
entity operating and updating the service. In aDNS, an example of trusted configuration is the registration policy of a zone.
Hence, updating the registration policy of a zone is considered a governance operation, which can only be initiated by {\em governance members},
and adopted by following a voting process directed by the {\em constitution} recorded at the start of the ledger. Each member is identified by a certificate, and each vote and decision
is recorded in the ledger to ensure that the governance decisions are transparent.

CCF governance also helps us solve the trust bootstrapping problem with aDNS. Since delegation policies are applied transitively, the governance
consortium of a root aDNS instance (at the apex of an attested zone) controls the baseline of what constitutes a valid
aDNS instance, and thus is particularly trusted. A possible model for aDNS deployment is for a non-profit organization to operate a TLD, in a
similar way that the non-profit Public Internet Registry operates the \name{.org} and~\name{.ngo} TLDs today. 
Alternatively, commercial entities such as CSPs may operate their own aDNS hierarchies, with their own root policies and aDNS implementations.

\mysubsection{Server features} 
Our aDNS implementation is focused on serving authoritative zones, therefore it cannot act as a recursive resolver.
It is capable of responding to requests over UDP, TCP (RFC7766), and HTTPS (RFC8484). However, our UDP interface always returns
a response with the truncation flag (TC) to force the client to re-try over TCP, to avoid having to implement protections against
DNS amplification and cache poisoning attacks.
We implement the resource record types required by aDNS, including DANE (RFC7671) and CAA (RFC6844).
We also introduce a new RR type for attestation (ATTEST), while policies are exposed as text (TXT).

We implement most DNSSEC features, including NSEC3. Expiration of RRsig records and rollover of ZSKs requires careful time management
which is challenging from an enclave application that can only access the (untrusted) host time.
Standard DNS records have a relative TTL, but RRsig records include both the start and end of validity of signatures as absolute timestamps.
To mitigate this, we introduce a notion of ledger time (inspired by blockchain techniques), so we can at least enforce monotonicity of host time.
It is also possible to monitor time in replicas to detect the clock of the primary server drifting too far. 
We assume a cron job on the host periodically calls aDNS to trigger the re-keying and key rollover endpoint.
If the host doesn't, then the signatures will expire and the service may become unavailable, but denial of service attack by the host is always possible anyway.

Our aDNS has a built-in ACME client, which can be configured either to use a private Pebble CA, or Let's Encrypt~\cite{letsencrypt}.
When an aDNS instance starts, we automatically provision certificates for each nodes, which are by convention labelled \name{ns0}..\name{ns}$N$,
where \name{ns0} is the primary and $N$ the number of replicas. 
These are used both for DoH and for the endpoints required by the aDNS protocol.

\cedric{recheck whether we use those names or a shared dns label}

\mysubsection{Endpoints} 
The table below summarizes the JSON RPC endpoints exposed by aDNS over HTTPS:
\vspace{-.7ex}
\begin{center}\scriptsize
  \begin{tabular}{@{}|l|l|r|}
  \hline
   Endpoint & Description \\
  \hline \hline
   \verb!/register_service! & Register a TEE to a name in the zone \\
   \verb!/register_delegation! & Delegate a sub-zone to a fresh aDNS instance \\
   \verb!/configure! & Configure a fresh aDNS instance 
   \\
   \verb!/get_certificate! & Get an ACME certificate for a registered name \\
   \verb!/resign! & Refresh RRsig records and rollover ZSK \\
   \verb!/acme_refresh! & Renew ACME certificates \\
   \verb!/endorsements! & Gets attestation evidence of this aDNS instance \\
  \hline
  \end{tabular}
\end{center}

Our implementation supports delegation and registration policies expressed as pure Javascript functions that return a boolean value. The functions have access to claims (key-value pairs) extracted from verified attestation reports. 

\subsection{Browser Extension}

To illustrate how Web browsers might support aDNS, we create a Firefox extension that implements the client protocol in \S\ref{sec:clientproto}.
We chose Firefox because, unlike Chromium-based browsers, it supports the {\tt getSecurityInfo()} API for inspecting the TLS connection (including the certificate) on a request.
We use the \name{WebRequest} API to intercept connections to aDNS websites, captured by the expression {\tt *://*.attested.name/*}. This assumes that all aDNS services fall under the \name{attested.name} suffix. In the future this may be replaced with a TLD, or there may be multiple subtrees of the DNS hierarchy with aDNS enabled.

We use the {\tt onBeforeRequest} event to send the initial DNS queries, and the {\tt onHeadersReceived} to validate the TLS connection.
We use either the Google or Cloudflare Trusted Recursive Resolver (TRR) to perform the additional DNS queries with DNS over HTTPS. These are available through {\tt browser.dns.resolve()} API (when the {\tt isTRR} flag is set in the response), but we prefer to query them directly with the Fetch API to ensure the Authentic Data (AD) flag is set, indicating successful DNSSEC validation.
It is also possible to use the extension without a TRR, but this configuration can only resolve A and AAAA requests. To work around this, we implement a mechanism that encodes ATTEST and TLSA records as a sequence of compressed AAAA records, e.g. when queried with the {\tt \_i.\_attest} prefix, where i is the fragment number.
Compression is important to ensure the fragments are properly cached. Our evaluation (see \S\ref{sub:evalserver}) shows that 5 fragments are sufficient for most reports, so our extension pre-flights 5 fragment queries in parallel.

We compile our Ravl attestation library to WebAssembly so that the attestation can be verified in parallel with the request, which only blocks when it reaches the {\tt onHeadersReceived} event.
Empirically, we observe that the verification of attestation reports and collaterals can complete in under 20ms, which is less than the connection latency for most websites. 
We compare the TLS certificate public key with the attested public key. If that check fails, we can interrupt the request and send the user to an error page.

For now, our extension discovers the registration policy and allows the user to inspect it using the extension's button, though we could allow the user to configure the policy instead.
This raises questions about how users would react to new UI signals about attested websites, which are left to future work.

\section{Evaluation}
\label{sec:eval}

\subsection{aDNS Server}
\label{sub:evalserver}

\begin{table}[t]
\centering \scriptsize
  \begin{tabular}{|l|l|r|}
  \hline
   C++ Class & Description &  LoC \\
  \hline \hline
  Ravl	& Remote attestation validation library & 7117 \\
  RFC1035 & Base DNS record types & 1278 \\
  RFC4034 & DNSSEC record types & 1353 \\
  RFC* & NSEC3, ETLS, DANE, CAA & 673 \\
  Resolver & DNSSEC implementation & 2472 \\
  CCFDNS & Endpoints (query, register...) & 3248 \\
  * & Serialization and utilities & 1146 \\
  \hline \hline
  CCF & Confidential Consortium Framework & $\sim$64K \\
  & CCF Dependencies (JSON, HTTP...) & $\sim$125K \\
  \hline \hline
  & Total Enclave TCB & $\sim$200K \\
  \hline
  \end{tabular}
  \vspace{-1em}
  \caption{TCB size of aDNS server}
  \vspace{-0.5em}
  \label{tab:tcb}
\end{table}

Our implementation of aDNS is coded in $\sim$17K lines of C++, whose breakdown is shown in Table~\ref{tab:tcb}.

\begin{figure}[t]
\centering
\begin{tikzpicture}
\datavisualization [scientific axes, visualize as smooth line, %
  x axis={min value=200, max value=1400, length=5.5cm, grid, label={Throughput (DNS queries/s)}}, %
 y axis={min value=18, max value=32,length=1.6cm, grid, ticks={step=5,minor steps between steps=5}, label={Latency (ms)}}]
data[headline={x, y}]
{
  250, 18.6
  500, 18.8
  1000, 20.84
  1330, 25.98
  1370, 31.2
};
\end{tikzpicture}
\vspace{-1em}
\caption{Throughput and latency of aDNS server}
\vspace{-1.3em}
  \label{fig:tl}
\end{figure}
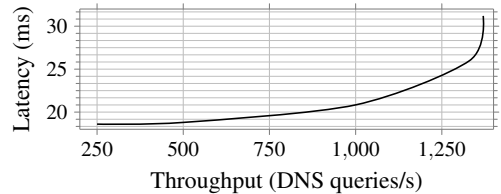

Figure~\ref{fig:tl} plots the throughput and latency of DNS queries using independent clients sending a single TCP request.
Our server runs on an Azure DCv2-series VM with 4 Xeon 8370C cores and 16GB of enclave protected memory (EPC), with
about 18ms round-trip time (RTT) to the client.
Although our server is limited to 1350 queries/s, this is sufficient in practice since we can add
any number of untrusted DNS caches in front of aDNS to scale up the volume of read requests.

\begin{table}[t]
\centering\scriptsize
  \begin{tabular}{|l|l|r|}
  \hline
   Endpoint & Context & Latency \\
  \hline \hline
   \verb!/register-service! & Register complete Intel SGX report & 263ms \\
   \verb!/register-service! & Register SGX, missing collaterals & 2.1s \\
   \verb!/register-service! & Register AMD SEV-SNP TEE & 218ms \\
   \verb!/get-certificate! & Get Let's Encrypt certificate & 72ms (+7.5s) \\
   \verb!/resign! & Re-sign zone with 454 records & 180ms \\
  \hline
  \end{tabular}
  \vspace{-1em}
  \caption{Non-DNS aDNS endpoint latency}
  \vspace{-0.5em}
  \label{tab:endplat}
\end{table}

Table~\ref{tab:endplat} shows the latency of non-DNS endpoints.
Incremental zone signing is not supported, so any record change causes all RRsig to be re-signed.
Because of NSEC3 and our fragmentation mechanism to encode ATTEST records into AAAA, our zones can be surprisingly large (up to 450 records per TEE registration), which explains the signing time.
Our implementation currently starts the ACME authorization process asynchronously as soon as a TEE is registered, which explains why the \verb`/get-certificate` endpoint is fast. 
It takes about 7 seconds for Let's Encrypt to return the certificate.

\begin{center}\scriptsize
\vspace{1mm}
  \begin{tabular}{|l|clc|c|}
  \hline
   Attestation report type & Full size & Compressed & Fragments \\
  \hline \hline
   Intel SGX & 5951 B & 2355 B & 5 \\
   AMD SEV-SNP & 3125 B & 630 B & 2 \\
  \hline
  \end{tabular}
\vspace{1mm}
\end{center}

Finally, we evaluate the size of ATTEST records. Since they are large and non-standard, they are not cached by most verifiers.
Hence, we compress and split them into fragmented AAAA record sets, to fit into 512 bytes UDP responses.
We find that all reports can fit in 5 compressed fragments.

\subsection{Browser Extension}

\begin{table}[t]
\centering\scriptsize
  \begin{tabular}{|l|c|c|c|c|}
  \hline
    & Default & aDNS & DoH & aDNS+DoH \\
  \hline \hline
  DNS (Address) & 1ms & 1ms & 43ms & 43ms \\
  DNS (TLSA/ATTEST) & - & 4ms & - & 51ms \\
  \hline
  TLS Handshake  & 40ms & 40ms & 40ms & 40ms \\
  Attestation (SGX)  & - & 27ms & - & 27ms  \\
  Attestation (SEV-SNP)  & - & 13ms & - & 13ms  \\
  \hline \hline
  Time to request & 41ms & 41ms & 83ms & 83ms \\
  \hline
  \end{tabular}
  \vspace{-1em}
  \caption{Connection latency with aDNS extension}
  \vspace{-1.5em}
  \label{tab:client}
\end{table}

Our Ravl library compiles to 2.9MB of WebAssembly and 130KB of JavaScript with Emscripten.
We also depend on pako for decompression of attestation records (46KB), a basic CBOR libraries (409 lines of JS) to interact with Ravl,
and jQuery for UI elements (86KB). The rest of the code is a simple 321 lines of JavaScript.
Table~\ref{tab:client} shows the total connection time to our inference service in 4 configurations: first using default browser settings and a local resolver,
second with the aDNS extension and the local resolver, using AAAA-encoded attestation and DANE records, and third using Cloudflare's
public DoH resolver.
Client-server RTT is $\sim$12ms.
Since all DNS queries are sent in parallel and the attestation validation is concurrent to the TLS handshake,
as long as the server RTT is over $\sim$10ms, there is no overhead at all for enabling aDNS.

\subsection{Sample Applications}
\label{sub:apps}
\noindent{\bf Inference Service.}
With the rising popularity of generative AI, concerns around data privacy, especially for the most performant commercial models that can only be queried through APIs, is harming adoption for sensitive scenarios.
Confidential computing is the most scalable solution to ensure confidentiality of prompts and models in cloud-based AI systems.

We deploy a Triton~\cite{tritonin57:online} container from Nvidia's container registry as a confidential Azure Container Instance (ACI) with 4 AMD EPYC 7763v cores and 16GB of memory.
We extend Ravl to support the additional Utility VM collateral of ACI attestation, used to prove that the OS image with the container runtime was built honestly by Azure.
We fork Azure's open source attestation sidecar so that it can register the underlying TEE to aDNS at startup.
We write an aDNS registration client in 367 lines of Go, which also obtains a Let's Encrypt certificate and runs an Nginx reverse proxy with the obtained certificate to access the application container service.
Including all configurations, our sidecar adds 674 lines of code and can be adapted to any containerized application.

\mysubsection{Privacy-Preserving Advertising}
The Privacy Sandbox~\cite{privacysandbox} is a set of browser technologies proposing to replace 3rd party cookies while providing advertisers with a privacy-preserving approach for targeted advertising and attribution measurement.
A key component of the Privacy Sandbox are Protected Audience services~\cite{protectedaudience}, which utilize confidential computing to host advertiser-provided bidding and auction logic in an attested sandboxed environment while protecting private interest-group information collected by the browser.
These services rely on a key management system (KMS) which releases private hybrid encryption keys only to known good TEEs.
Since Azure's implementation of the Privacy Sandbox KMS~\cite{azurekms} is a CCF application, we modify the built-in ACME client of CCF to register the nodes and get certificates from aDNS instead.
Our patch uses 414 lines of C++.

\section{Related Work \& Conclusion}
\label{sec:related}

The state of the art is to verify attestation either during or after the TLS handshake using remotely attested TLS (RA-TLS).
Many RA-TLS variants have been proposed~\cite{tlsrabind, knauth2018integrating, niemi2021trusted, hamidy2023tc4se} (including a draft standard~\cite{fossati-tls-attestation-08})
and implemented (e.g. in the Intel SGX SDK and libraries~\cite{ratstls}).
Sardar et al.~\cite{10752524} built a formal model and identified vulnerabilities in Intel's protocol~\cite{knauth2018integrating}.
All variants of RA-TLS share the downside of breaking compatibility with legacy client, either because they change TLS or use self-signed certificates.
In contrast, since aDNS operates below the transport and application layers, it is compatible with all clients
(their networking stack may be enlightened e.g. by adapting their DNS resolver) and it is more efficient by leveraging distributed DNS caches and allowing concurrent attestation validation and TLS handshake. 

Another related line of works focuses on delegating attestation verification to trusted third parties called attestation services (AS). AS operated by
cloud providers (e.g. Azure~\cite{maa}) and hardware vendors (Intel, Nvidia, Arm) typically implement the Entity Attestation Token~\cite{ietf-rats-eat-31}
standard, that is, they consume attestation reports and produce JWT tokens summarizing the contents of the report and the additional policies applied by the AS.
The client's (or "relying party") trust in the AS (or "verifier") is a major concern---some AS run in TEEs but this still requires clients to understand attestation.
It may seem logical to depend on a TEE vendor's AS (given TEE security assumes the vendor is honest). However, enforcing service and cloud specific policies in a global AS is challenging,
and depending on them further creates latency, availability and scalability issues 
%
aDNS can be described as a distributed, transparent, attested AS that can provide compatibility with EAT relying parties
(as explained in \S\ref{sec:eat}) while solving these issues.  
Akama et al.~\cite{akama2024rawebsremoteattestationweb} propose a way to protect legacy Web clients of a confidential service using certificate transparency (CT)
to detect conflicting attested and non-attested service certificates;
aDNS goes further by preventing certificate issuance to non-TEE entities by ACME-compatible CAs, though it also relies on CT to detect mis-issuance by other CAs.

In the past decade, many TEE-based applications have been proposed for Web search~\cite{pires2018cyclosa, mokhtar2017x},
online payments~\cite{li2020privacy,lind2019teechain}, video streaming~\cite{da2019privatube}, file sharing~\cite{fuhry2020segshare}, password management~\cite{krawiecka2018safekeeper,liang2017man}, and many others domains.
Protecting data and model confidentiality of AI systems is by far the most popular type of applications~\cite{vaswani2023confidential,tramer2018slalom,mo2020darknetz, quoc2020securetf, zhang2021citadel, yuhala2021plinius,mondal2021poster,mo2019efficient}.
Paju et al. conducted a survey of 223 TEE applications~\cite{paju2023sok}, most of which rely on a custom client.
%
%
However, as more applications appear, this approach cannot scale.
aDNS is the first service attestation architecture to provide low connection overhead, legacy client compatibility and meaningful security properties for all clients in the face of various compromise scenarios.

\clearpage 

\iffull 
\small 
\bibliographystyle{plain}
\bibliography{\jobname}

\begin{thebibliography}{10}

\bibitem{letsencrypt}
Josh Aas, Richard Barnes, Benton Case, Zakir Durumeric, Peter Eckersley, Alan
  Flores-L\'{o}pez, J.~Alex Halderman, Jacob Hoffman-Andrews, James Kasten,
  Eric Rescorla, Seth Schoen, and Brad Warren.
\newblock Let's encrypt: An automated certificate authority to encrypt the
  entire web.
\newblock In {\em Proceedings of the 2019 ACM SIGSAC Conference on Computer and
  Communications Security}, CCS '19, page 2473–2487, New York, NY, USA, 2019.
  Association for Computing Machinery.

\bibitem{akama2024rawebsremoteattestationweb}
Kosei Akama, Yoshimichi Nakatsuka, Korry Luke, Masaaki Sato, and Keisuke
  Uehara.
\newblock Ra-webs: Remote attestation for web services, 2024.

\bibitem{contour18}
Mustafa Al-Bassam and Sarah Meiklejohn.
\newblock Contour: A practical system for binary transparency.
\newblock In {\em Data Privacy Management, Cryptocurrencies and Blockchain
  Technology}, pages 94--110, 2018.

\bibitem{sev2020strengthening}
{AMD}.
\newblock Strengthening {VM} isolation with integrity protection and more.
\newblock {\em White Paper, January}, 53:1450--1465, 2020.

\bibitem{rfc8555}
Richard Barnes, Jacob Hoffman-Andrews, Daniel McCarney, and James Kasten.
\newblock {Automatic Certificate Management Environment (ACME)}.
\newblock RFC 8555, March 2019.

\bibitem{beugin2023interestdisclosing}
Yohan Beugin and Patrick McDaniel.
\newblock Interest-disclosing mechanisms for advertising are privacy-exposing
  (not preserving), 2023.

\bibitem{cab}
{CA/Browser Forum}.
\newblock Baseline requirements for {TLS} server certificates.
\newblock
  \url{https://cabforum.org/working-groups/server/baseline-requirements/documents/},
  2024.

\bibitem{cheng2023intel}
Pau-Chen Cheng, Wojciech Ozga, Enriquillo Valdez, Salman Ahmed, Zhongshu Gu,
  Hani Jamjoom, Hubertus Franke, and James Bottomley.
\newblock Intel {TDX} demystified: A top-down approach, 2023.

\bibitem{signalrecovery}
Graeme Connell, Vivian Fang, Rolfe Schmidt, Emma Dauterman, and Raluca~Ada
  Popa.
\newblock Secret key recovery in a {Global-Scale} {End-to-End} encryption
  system.
\newblock In {\em 18th USENIX Symposium on Operating Systems Design and
  Implementation (OSDI 24)}, pages 703--719, Santa Clara, CA, July 2024. USENIX
  Association.

\bibitem{da2019privatube}
Simon Da~Silva, Sonia Ben~Mokhtar, Stefan Contiu, Daniel N{\'e}gru, Laurent
  R{\'e}veill{\`e}re, and Etienne Rivi{\`e}re.
\newblock Privatube: Privacy-preserving edge-assisted video streaming.
\newblock In {\em Proceedings of the 20th International Middleware Conference},
  pages 189--201, 2019.

\bibitem{delignat2023should}
Antoine Delignat-Lavaud, C{\'e}dric Fournet, Kapil Vaswani, Sylvan Clebsch,
  Maik Riechert, Manuel Costa, and Mark Russinovich.
\newblock Why should i trust your code?
\newblock {\em Communications of the ACM}, 67(1):68--76, 2023.

\bibitem{dhanuskodi2023creating}
Gobikrishna Dhanuskodi, Sudeshna Guha, Vidhya Krishnan, Aruna Manjunatha,
  Michael O'Connor, Rob Nertney, and Phil Rogers.
\newblock Creating the first confidential gpus: The team at nvidia brings
  confidentiality and integrity to user code and data for accelerated
  computing.
\newblock {\em Queue}, 21(4):68--93, 2023.

\bibitem{ratstls}
Zia~Zhang et~al.
\newblock {RATS} architecture based {TLS} using librats.
\newblock \url{ https://github.com/inclavare-containers/rats-tls}, 2024.

\bibitem{fuhry2020segshare}
Benny Fuhry, Lina Hirschoff, Samuel Koesnadi, and Florian Kerschbaum.
\newblock {SeGShare}: Secure group file sharing in the cloud using enclaves.
\newblock In {\em 2020 50th Annual IEEE/IFIP International Conference on
  Dependable Systems and Networks (DSN)}, pages 476--488. IEEE, 2020.

\bibitem{tlsrabind}
Kenneth Goldman, Ronald Perez, and Reiner Sailer.
\newblock Linking remote attestation to secure tunnel endpoints.
\newblock In {\em Proceedings of the First ACM Workshop on Scalable Trusted
  Computing}, STC '06, page 21–24, New York, NY, USA, 2006. Association for
  Computing Machinery.

\bibitem{protectedaudience}
Google.
\newblock Protected audience api overview.
\newblock
  \url{https://developers.google.com/privacy-sandbox/relevance/protected-audience}.

\bibitem{privacysandbox}
Google.
\newblock What is the privacy sandbox?
\newblock \url{https://developers.google.com/privacy-sandbox/overview}, 2023.

\bibitem{hamidy2023tc4se}
Gilang~Mentari Hamidy, Sri Yulianti, Pieter Philippaerts, and Wouter Joosen.
\newblock Tc4se: A high-performance trusted channel mechanism for secure
  enclave-based trusted execution environments.
\newblock In {\em International Conference on Information Security}, pages
  246--264. Springer, 2023.

\bibitem{rfc6698}
Paul~E. Hoffman and Jakob Schlyter.
\newblock {The DNS-Based Authentication of Named Entities (DANE) Transport
  Layer Security (TLS) Protocol: TLSA}.
\newblock RFC 6698, August 2012.

\bibitem{sgx}
Intel.
\newblock Software guard extensions.
\newblock \url{https://software.intel.com/en-us/sgx} (Accessed on 12/13/2019).

\bibitem{signalcontacts}
Daniel Kales, Christian Rechberger, Thomas Schneider, Matthias Senker, and
  Christian Weinert.
\newblock Mobile private contact discovery at scale.
\newblock In {\em 28th USENIX Security Symposium (USENIX Security 19)}, pages
  1447--1464, Santa Clara, CA, August 2019. USENIX Association.

\bibitem{kalodner2015empirical}
Harry~A Kalodner, Miles Carlsten, Paul~M Ellenbogen, Joseph Bonneau, and Arvind
  Narayanan.
\newblock An empirical study of namecoin and lessons for decentralized
  namespace design.
\newblock In {\em WEIS}, volume~1, pages 1--23, 2015.

\bibitem{knauth2018integrating}
Thomas Knauth, Michael Steiner, Somnath Chakrabarti, Li~Lei, Cedric Xing, and
  Mona Vij.
\newblock Integrating remote attestation with transport layer security.
\newblock {\em arXiv preprint arXiv:1801.05863}, 2018.

\bibitem{krawiecka2018safekeeper}
Klaudia Krawiecka, Arseny Kurnikov, Andrew Paverd, Mohammad Mannan, and
  N~Asokan.
\newblock {Safekeeper}: Protecting {Web} passwords using trusted execution
  environments.
\newblock In {\em Proceedings of the 2018 World Wide Web Conference}, pages
  349--358, 2018.

\bibitem{chromedane}
Adam Langley.
\newblock Why not dane in browsers?
\newblock \url{https://www.imperialviolet.org/2015/01/17/notdane.html}, 2015.

\bibitem{certificate-transparency-2014}
Ben Laurie.
\newblock Certificate transparency.
\newblock {\em Commun. ACM}, pages 40--46, sep 2014.

\bibitem{li2020privacy}
Peng Li, Xiaofei Luo, Toshiaki Miyazaki, and Song Guo.
\newblock Privacy-preserving payment channel networks using trusted execution
  environment.
\newblock In {\em ICC 2020-2020 IEEE International Conference on Communications
  (ICC)}, pages 1--6. IEEE, 2020.

\bibitem{8705911}
Wubin Li, Yves Lemieux, Jing Gao, Zhuofeng Zhao, and Yanbo Han.
\newblock Service mesh: Challenges, state of the art, and future research
  opportunities.
\newblock In {\em 2019 IEEE International Conference on Service-Oriented System
  Engineering (SOSE)}, pages 122--1225, 2019.

\bibitem{cca}
Xupeng Li, Xuheng Li, Christoffer Dall, Ronghui Gu, Jason Nieh, Yousuf Sait,
  and Gareth Stockwell.
\newblock Design and verification of the arm confidential compute architecture.
\newblock In {\em 16th USENIX Symposium on Operating Systems Design and
  Implementation (OSDI 22)}, pages 465--484, Carlsbad, CA, July 2022. USENIX
  Association.

\bibitem{lian2013measuring}
Wilson Lian, Eric Rescorla, Hovav Shacham, and Stefan Savage.
\newblock Measuring the practical impact of $\{$DNSSEC$\}$ deployment.
\newblock In {\em 22nd USENIX Security Symposium (USENIX Security 13)}, pages
  573--588, 2013.

\bibitem{liang2017man}
Xueping Liang, Sachin Shetty, Lingchen Zhang, Charles Kamhoua, and Kevin Kwiat.
\newblock Man in the cloud ({MITC}) defender: {SGX}-based user credential
  protection for synchronization applications in cloud computing platform.
\newblock In {\em 2017 IEEE 10th International Conference on Cloud Computing
  (CLOUD)}, pages 302--309. IEEE, 2017.

\bibitem{lind2019teechain}
Joshua Lind, Oded Naor, Ittay Eyal, Florian Kelbert, Emin~G{\"u}n Sirer, and
  Peter Pietzuch.
\newblock Teechain: a secure payment network with asynchronous blockchain
  access.
\newblock In {\em Proceedings of the 27th ACM Symposium on Operating Systems
  Principles}, pages 63--79, 2019.

\bibitem{ietf-rats-eat-31}
Laurence Lundblade, Giridhar Mandyam, Jeremy O'Donoghue, and Carl Wallace.
\newblock {The Entity Attestation Token (EAT)}.
\newblock Internet-Draft draft-ietf-rats-eat-31, Internet Engineering Task
  Force, September 2024.
\newblock Work in Progress.

\bibitem{ietf-rats-eat-25}
Laurence Lundblade, Giridhar Mandyam, Jeremy O'Donoghue, and Carl Wallace.
\newblock {The Entity Attestation Token (EAT)}.
\newblock Internet-Draft draft-ietf-rats-eat-25, Internet Engineering Task
  Force, January 2024.
\newblock Work in Progress.

\bibitem{envoy}
Lyft.
\newblock {Envoy} proxy.
\newblock \url{https://www.envoyproxy.io/}, 2024.

\bibitem{signal}
Moxie Marlinspike.
\newblock Technology preview: Private contact discovery for {Signal}.
\newblock \url{https://signal.org/blog/private-contact-discovery/}.

\bibitem{maa}
Microsoft.
\newblock Azure attestation.
\newblock \url{https://azure.microsoft.com/en-us/products/azure-attestation/},
  2022.

\bibitem{azurekms}
Microsoft.
\newblock Azure privacy sandbox kms.
\newblock \url{https://github.com/microsoft/azure-privacy-sandbox-kms}, 2024.

\bibitem{mo2019efficient}
Fan Mo and Hamed Haddadi.
\newblock Efficient and private federated learning using {TEE}.
\newblock In {\em Proc. EuroSys Conf., Dresden, Germany}, 2019.

\bibitem{mo2020darknetz}
Fan Mo, Ali~Shahin Shamsabadi, Kleomenis Katevas, Soteris Demetriou, Ilias
  Leontiadis, Andrea Cavallaro, and Hamed Haddadi.
\newblock Darknetz: towards model privacy at the edge using trusted execution
  environments.
\newblock In {\em Proceedings of the 18th International Conference on Mobile
  Systems, Applications, and Services}, pages 161--174, 2020.

\bibitem{mokhtar2017x}
Sonia~Ben Mokhtar, Antoine Boutet, Pascal Felber, Marcelo Pasin, Rafael Pires,
  and Valerio Schiavoni.
\newblock X-search: revisiting private web search using intel sgx.
\newblock In {\em Proceedings of the 18th ACM/IFIP/USENIX Middleware
  Conference}, pages 198--208, 2017.

\bibitem{mondal2021poster}
Arup Mondal, Yash More, Ruthu~Hulikal Rooparaghunath, and Debayan Gupta.
\newblock Flatee: Federated learning across trusted execution environments.
\newblock In {\em 2021 IEEE European Symposium on Security and Privacy
  (EuroS\&P)}, pages 707--709. IEEE, 2021.

\bibitem{niemi2021trusted}
Arto Niemi, Vasile Adrian~Bogdan Pop, and Jan-Erik Ekberg.
\newblock Trusted sockets layer: A tls 1.3 based trusted channel protocol.
\newblock In {\em Nordic Conference on Secure IT Systems}, pages 175--191.
  Springer, 2021.

\bibitem{nottingham2021playing}
Mark Nottingham.
\newblock Playing fair in the privacy sandbox: Competition, privacy and
  interoperability standards.
\newblock {\em Privacy and Interoperability Standards (February 3, 2021)},
  2021.

\bibitem{ongaro2014search}
Diego Ongaro and John Ousterhout.
\newblock In search of an understandable consensus algorithm.
\newblock In {\em 2014 USENIX annual technical conference (USENIX ATC 14)},
  pages 305--319, 2014.

\bibitem{osterweil2008quantifying}
Eric Osterweil, Michael Ryan, Dan Massey, and Lixia Zhang.
\newblock Quantifying the operational status of the {DNSSEC} deployment.
\newblock In {\em Proceedings of the 8th ACM SIGCOMM conference on Internet
  measurement}, pages 231--242, 2008.

\bibitem{paju2023sok}
Arttu Paju, Muhammad~Owais Javed, Juha Nurmi, Juha Savim{\"a}ki, Brian
  McGillion, and Billy~Bob Brumley.
\newblock {SoK}: A systematic review of tee usage for developing trusted
  applications.
\newblock In {\em Proceedings of the 18th International Conference on
  Availability, Reliability and Security}, pages 1--15, 2023.

\bibitem{pires2018cyclosa}
Rafael Pires, David Goltzsche, Sonia~Ben Mokhtar, Sara Bouchenak, Antoine
  Boutet, Pascal Felber, R{\"u}diger Kapitza, Marcelo Pasin, and Valerio
  Schiavoni.
\newblock {CYCLOSA}: Decentralizing private web search through {SGX}-based
  browser extensions.
\newblock In {\em 2018 IEEE 38th International Conference on Distributed
  Computing Systems (ICDCS)}, pages 467--477. IEEE, 2018.

\bibitem{quoc2020securetf}
Do~Le Quoc, Franz Gregor, Sergei Arnautov, Roland Kunkel, Pramod Bhatotia, and
  Christof Fetzer.
\newblock Securetf: A secure tensorflow framework.
\newblock In {\em Proceedings of the 21st International Middleware Conference},
  pages 44--59, 2020.

\bibitem{russinovich2019ccf}
Mark Russinovich, Edward Ashton, Christine Avanessians, Miguel Castro, Amaury
  Chamayou, Sylvan Clebsch, Manuel Costa, C{\'e}dric Fournet, Matthew Kerner,
  Sid Krishna, et~al.
\newblock Ccf: A framework for building confidential verifiable replicated
  services.
\newblock {\em Microsoft, Redmond, WA, USA, Tech. Rep. MSR-TR-2019-16}, 2019.

\bibitem{oidcdisco}
N.~Sakimura, J.~Bradley, M.~Jones, and E.~Jay.
\newblock {OpenID Connect} discovery 1.0, December 2023.

\bibitem{10752524}
Muhammad~Usama Sardar, Arto Niemi, Hannes Tschofenig, and Thomas Fossati.
\newblock Towards validation of tls 1.3 formal model and vulnerabilities in
  intel’s ra-tls protocol.
\newblock {\em IEEE Access}, 12:173670--173685, 2024.

\bibitem{rfc9460}
Benjamin~M. Schwartz, Mike Bishop, and Erik Nygren.
\newblock {Service Binding and Parameter Specification via the DNS (SVCB and
  HTTPS Resource Records)}.
\newblock RFC 9460, November 2023.

\bibitem{shamis2021pac}
Alex Shamis, Peter Pietzuch, Miguel Castro, Cedric Fournet, Edward Ashton,
  Amaury Chamayou, Sylvan Clebsch, Antoine Delignat-Lavaud, Matthew Kerner,
  Julien Maffre, Olga Vrousgou, Christoph~M. Wintersteiger, Manuel Costa, and
  Mark Russinovich.
\newblock {IA-CCF}: Individual accountability for permissioned ledgers.
\newblock In {\em 19th USENIX Symposium on Networked Systems Design and
  Implementation (NSDI)}, pages 467--491, 2022.

\bibitem{szurdi2014long}
Janos Szurdi, Balazs Kocso, Gabor Cseh, Jonathan Spring, Mark Felegyhazi, and
  Chris Kanich.
\newblock The long “taile” of typosquatting domain names.
\newblock In {\em 23rd USENIX Security Symposium (USENIX Security 14)}, pages
  191--206, 2014.

\bibitem{tramer2018slalom}
Florian Tramer and Dan Boneh.
\newblock Slalom: Fast, verifiable and private execution of neural networks in
  trusted hardware.
\newblock {\em arXiv preprint arXiv:1806.03287}, 2018.

\bibitem{tritonin57:online}
triton-inference-server/server: The triton inference server provides an
  optimized cloud and edge inferencing solution.
\newblock \url{https://github.com/triton-inference-server/server}.
\newblock (Accessed on 11/04/2021).

\bibitem{fossati-tls-attestation-08}
Hannes Tschofenig, Yaron Sheffer, Paul Howard, Ionuț Mihalcea, Yogesh
  Deshpande, Arto Niemi, and Thomas Fossati.
\newblock {Using Attestation in Transport Layer Security (TLS) and Datagram
  Transport Layer Security (DTLS)}.
\newblock Internet-Draft draft-fossati-tls-attestation-08, Internet Engineering
  Task Force, October 2024.
\newblock Work in Progress.

\bibitem{secspider}
George~Mason University.
\newblock Global dnssec deployment tracking.
\newblock \url{https://secspider.net/}, 2024.

\bibitem{vaswani2023confidential}
Kapil Vaswani, Stavros Volos, C{\'e}dric Fournet, Antonio~Nino Diaz, Ken
  Gordon, Balaji Vembu, Sam Webster, David Chisnall, Saurabh Kulkarni, Graham
  Cunningham, et~al.
\newblock Confidential computing within an {AI} accelerator.
\newblock In {\em 2023 USENIX Annual Technical Conference (USENIX ATC 23)},
  pages 501--518, 2023.

\bibitem{rfc7250}
Paul Wouters, Hannes Tschofenig, John~IETF Gilmore, Samuel Weiler, and Tero
  Kivinen.
\newblock {Using Raw Public Keys in Transport Layer Security (TLS) and Datagram
  Transport Layer Security (DTLS)}.
\newblock RFC 7250, June 2014.

\bibitem{yuhala2021plinius}
Peterson Yuhala, Pascal Felber, Valerio Schiavoni, and Alain Tchana.
\newblock Plinius: Secure and persistent machine learning model training.
\newblock In {\em 2021 51st Annual IEEE/IFIP International Conference on
  Dependable Systems and Networks (DSN)}, pages 52--62. IEEE, 2021.

\bibitem{zhang2021citadel}
Chengliang Zhang, Junzhe Xia, Baichen Yang, Huancheng Puyang, Wei Wang,
  Ruichuan Chen, Istemi~Ekin Akkus, Paarijaat Aditya, and Feng Yan.
\newblock Citadel: Protecting data privacy and model confidentiality for
  collaborative learning.
\newblock In {\em Proceedings of the ACM Symposium on Cloud Computing}, pages
  546--561, 2021.

\end{thebibliography}

\normalsize

\fi 

\appendix

\section*{Appendix}

We provide additional materials, including a description of a sample confidential service 
consisting of multiple TEEs and of the corresponding TEE registration policies for this service (\S\ref{sec:l7}); 
an aDNS extension issuing EAT tokens based on registered verified attestations to relying parties (\S\ref{sec:eat});
an overview of a confidential service for the Privacy Sandbox (\S\ref{sec:sandbox}); and an extended discussion 
of attested DNS vs attested CAs (\S\ref{sec:ca}).

\section{Sample service: scalable L7 load balancing}\label{sec:l7}


To illustrate aDNS-enforced policies for registration and certificate issuance, let us consider a distributed service with two roles assigned to TEEs (see also Figure~\ref{fig:architecture}):
\begin{enumerate}
\item front-end servers that terminate connections to a shared URL \name{https://www.s.conf}, 
deal with client authentication and authorization, and dispatch their tasks to the back-end;  
\item back-end servers that accepts tasks from the front-end.
\end{enumerate}

Each front-end TEE creates an X.509 key to accept client requests over TLS, and gets its endorsed by the CA, as explained above. 
It also has a DANE key, used for client authentication with the back-end TEEs. 
Each back-end TEE creates a single DANE key to accept connection requests from front-ends.
This limits the need for trusted Web certificates only to client-facing frontend TEEs.

The aDNS instance that controls the service registers both kinds of TEEs, 
and its policy (sample shown below) ensures they use pairwise-distinct names of the form \name{node}$n$\name{.front-end.s.}\apex and 
\name{node}$n$\name{.back-end.s.}\apex{} with code and configuration that match their role. 

\begin{lstlisting}
function validate(attestation, config) {
  const role = config["role"];
  const values = {
      "front-end": {
          hostdata: 0xDEADC0DE,
          measurement: 0xFEEDFACE
      },
      "back-end": {
          hostdata: 0xBAADF00D,
          measurement: 0x8BADF00D
      }
  };

  return role in values &&
      attestation["hostdata"] === values[role].hostdata &&
      attestation["measurement"] === values[role].measurement;
}
\end{lstlisting}

Each front-end regularly queries aDNS to get the names and DANE keys of every active back-end as part of their TLSA records.
Symmetrically, each back-end caches the names and DANE keys of every active front-end. 
Hence, whenever a front-end opens a TLS connection to a back-end, both parties can present their names and raw keys in their ClientCertificate and ServerCertificate messages.

As long as they cache live TLSA records for these keys, they are guaranteed that their peers have been transparently attested and authorized for their role. 
They may simply rely on their recent verification by aDNS at registration time or, for additional assurance, fetch and verify the other party's attestation record.
The latter approach may be useful, for instance, to confirm that the versions of the code at the two endpoint are compatible, 
or to ensure that the attested code of the front-end
independently verifies the attestation of the front-end before trusting it with the client's data. 

\section{A Discoverable, Transparent Entity Attestation Service}\label{sec:eat}

\paragraph{Entity Attestation Tokens (review)}
Attestation services enable clients to delegate the verification of attestation evidence to a trusted third party. 
They take as input attestation reports and product a special type of JSON Web Token called an Entity Attestation Tokens (EAT)~\cite{ietf-rats-eat-25}, which
encodes the properties of the TEE into a set of platform-agnostic claims.
These claims can be be used as input of declarative authentication and authorization policies, while providing an abstract layer atop the underlying,
hardware-specific attestation evidence and their cumbersome collaterals.
Although these services are convenient, they grow the TCB and reduce auditability, as clients only authenticate tokens.

\paragraph{OpenID Discovery (review)}
The OpenID Connect Discovery~\cite{oidcdisco} protocol is used by token issuers to advertise their capabilities to clients at a fixed, well-known URL, thereby enabling 
clients to dynamically obtain up-to-date information about the issuer's configuration, including the public keys used to sign ID tokens.
This can similarly lead to a significant TCB extension, as the compromise of the discovery Web server (or its TLS keys) can lead to the compromise of its discoverable services. 

\paragraph{EAT as an auxiliary aDNS API} 
Our main goal is to issue standard-compliant EAT tokens that mirror the verified, transparent attestation evidence recorded by aDNS for the services in its zone, for the benefit of relying parties that consume EATs instead of directly verifying attestation evidence. 
Clearly, fetching attestation from aDNS scales better than clients fetching an EAT token from a centralized attestation endpoint, but this is still a useful
mechanism to support applications built on EAT.
%
%
We ensure that the EAT issuance service has a minimal, well-defined TCB; that it strictly manages the TTL of the EATs it issues as part of the registration policies it enforces; 
and that it automatically manages the keys for issuance and for discovery. 

\paragraph{Implementation}
Our EAT service is directly built into aDNS using about 200 lines of C++. When queried for a token for a service name, it fetches the latest unexpired
attestation from its zone configuration, and translates the Ravl properties into EAT claims.
To this end, we introduce the following new endpoints:
\begin{itemize} 
\item \url{/common/v2.0/.well-known/openid-configuration} returns the configuration of the EAT discovery service, notably the discovery and token issuance URL;
\item \url{/common/discovery/v2.0/keys} returns the JWKS of all public keys to verify recent and future EATs. 
\item \url{/common/oauth2/v2.0/token?service={S}} issues a token for the TEE(s) registered at name S.
\item \url{/create-signing-key} samples a fresh keypair for token issuance, starts advertising the jwk for the public key, and keeps the private key within the CCF TEEs. This enables the (authenticated) service operator to trigger rotations without exposing private keys.  
\end{itemize}
To prevent any confusion with core aDNS functionality, the EAT endpoint of aDNS can be served from a separate interface, using an independent certificate from the nameserver's primary name.
Getting an EAT token for the inference service takes under 78ms (locally) and 160ms (remotely, with a RTT of about 15ms), which is much faster
than any public attestation service we tested. 

\section{Attested DNS vs Attested CA}\label{sec:ca}

The reader may wonder why we decided to build a transparent attested DNS, rather than a transparent attested certification authority.
Several open source confidential applications embed their attestation reports in their (self-signed) certificate, using X.509 extensions.
We could extend the ACME protocol so that the CA can verify attestation from the TEE CSR and record it in the certificate. 
While this requires some difficult standardization work at the IETF and CA/Browser forum, we think in the long term this approach
is inferior to attested DNS for the following reasons:

\begin{itemize}

\item For most humans, domain names are the only meaningful part of Web certificates, and their only part surfaced in modern UIs. 

\item Embedding auxiliary evidence and attestation reports within certificates is awkward. This yields large certificate chains that may slow down connection establishment, or even break some networking stacks. In contrast, DNS is great at communicating and caching these materials and their validation ahead of connection establishment. 

\item Similarly, DNS features tighter time-based policies, 
on the scale of global network RTTs, whereas certificates remain valid for many years. 

\item On the client-side, validating this evidence in callbacks while processing certificate chains involves customizing the transport level, often shared between many applications. It may also be hard to delay the use of the connection (including 0RTT traffic, cookies etc) until 
the attested evidence has been checked. 

\item Deploying a new CA comes with high risks and overheads, inasmuch
as root CAs are heavily regulated and the scope of the certificates they endorse is hard to bound. 
In contrast, the controlled delegation of trust to subdomains is a central feature of DNS, so that we can deploy our proposed "confidential zone" without interfering with other domains. 

\item Together with certificate transparency and ACME, let's encrypt already provides an independent, trustworthy CA with strong practical security. This limits the value of a fresh, TEE-based implementation. 
Our approach supplements their security with attested DNS security. 

\end{itemize}

\section{Privacy Sandbox}\label{sec:sandbox}
Mozilla, Apple and Google have publicly committed to removing support for third party cookies in Firefox, Safari and Chrome by the end of 2024. This change is disruptive to the online advertising industry, which relies on third-party cookies to track users across websites in order to serve targeted ads (a feature that has been widely abused for many unauthorized purposes).
In response, Google has introduced a new set of features in Chrome called the Privacy Sandbox designed to enable targeted advertising while limiting the ability for advertisers to track users.
A key component of the Privacy Sandbox are Protected Audience services, which utilize confidential computing to run an auction between advertisers for each ad space based on the user's interest groups maintained by the browser. Interest groups are registered when visiting websites; to display an ad, the browser encrypts the list of interest groups under a public key whose associated private key can only be released to an attested TEE running the approved code.
Advertisers participate in the auction by providing bidding scripts that can locally output a bid based on the interest groups, and some generic context (e.g. the user's location or language). The bidding and auction sandbox is meant to prevent any of this information from leaking to the advertiser.
While Privacy Sandbox has been criticized both for its ineffectiveness at protecting privacy~\cite{beugin2023interestdisclosing} as well as for the conflict of interest with Google's advertising business~\cite{nottingham2021playing}, we consider it a critical application irrespective of its technical merits because of its potential to become the most widely deployed confidential computing Web service.

In Google's current implementation, the key used to encrypt interest groups is generated in a TEE, before being split into two shares managed by independent organizations (called the {\em coordinators}). The threat model is that at least one of the coordinators is trusted, and if the coordinators decide to collude (e.g., by giving their share to each other, or by accepting a key not created by the key generation TEE), both of them can decrypt all user data without any risk of being blamed.
We observe that this risk can be significantly reduced if instead of fetching the interest group encryption key from an untrusted HTTPS endpoint at the primary coordinator, this key was instead registered by the key generation TEE on aDNS and distributed through DANE records to browsers. Such an approach has the following benefits:
\begin{itemize}
\item the attested DNS service verifies and stores the attestation of the key generation TEE, so users can audit the generation of the key;
\item the aDNS ledger makes the key transparent and it is no longer possible for the primary coordinator to do a targeted attack against a class of users;
\item it separates the generation and distribution of the key, so that different browser vendors (e.g. Firefox, Edge) can distribute the key for Google's bidding and auction services while enforcing their own policies;
\item it reuses the distributed caching infrastructure of DNS.
\end{itemize}

\end{document}